\documentclass[aps,pra,10pt,superscriptaddress,twocolumn,amsmath,amssymb,amsthm
]{revtex4-2}

\usepackage[dvipsnames]{xcolor}
\usepackage{soul}
\usepackage{physics}

\definecolor{revdcolor}{HTML}{B45309}

\usepackage{graphicx}
\usepackage{dcolumn}
\usepackage{bm}
\usepackage[utf8]{inputenc}

\usepackage{braket}
\usepackage{physics}
\usepackage{mathtools}
\usepackage{tikz-cd}
\usepackage{dsfont}
\usepackage{scalerel}
\usepackage{cancel}
\usepackage{mathbbol}
\DeclareSymbolFontAlphabet{\amsmathbb}{AMSb}%

\newcommand{\beq}{\begin{equation}}
\newcommand{\eeq}{\end{equation}}
\newcommand{\bse}{\begin{subequations}}
\newcommand{\ese}{\end{subequations}}
\newcommand{\bea}{\begin{eqnarray}}
\newcommand{\eea}{\end{eqnarray}}
\newcommand{\tp}{\intercal}
\def\no{n_1}
\def\nt{n_2}
\def\mV{\mathbb{V}}
\def\mVb{{\mathbb{V}}}

\def\mJb{{\mathsf{J}}}
\def\mS{\mathsf{S}}
\def\mSb{{\mathsf{S}}}

\def\mOb{{\mathsf{O}}}
\def\mObp{{\mathsf{O}}^\prime}
\def\mI1{\mathbb{1}_{2\times 2}}

\def\mSigb{{\mathsf{\Sigma}}}
\def\mOmb{{\mathsf{\Omega}}}
\def\no{{\rm n}_1}
\def\nt{{\rm n}_2}
\def\hR{\hat{\mathbb{R}}}

\def\hUS{\hat{U}_{\mathsf{S}}}
\def\hUSd{\hat{U}_{\mathsf{S_d}}}
\def\hUO{\hat{U}_{\mathsf{O}}}
\def\hUOp{\hat{U}_{\mathsf{O^\prime}}}

\def\bx{{\bf x}}
\def\bxo{{\bf x}_1}
\def\bxt{{\bf x}_2}
\def\hrho{\hat{\rho}}
\def\ie{{\it i.e}, }
\def\diag{\text{diag}}
\def\calI{{\cal I}}

\begin{document}


\title{Twisted Gaussian Schell States in Quantum Optics: Twist-Assisted Nonclassicality and Entanglement}

\author{Fabricio Toscano} 
\email{toscano@if.ufrj.br \ (corresponding author) }
\affiliation{Instituto de F\'isica, Universidade Federal do Rio de Janeiro,
  21941-972, Rio de Janeiro, Brazil}
\author{G.~Ca\~{n}as}
\affiliation{Departamento de F\'isica, Universidad del B\'io-B\'io, Collao 1202, 5-C Concepci\'on, Chile}

\author{A. Z. Khoury}
\affiliation{Instituto de F\'{i}sica, Universidade Federal Fluminense, 24210-346 Niter\'{o}i, RJ, Brazil}

\author{P. H. Souto Ribeiro}
\affiliation{Departamento de F\'{i}sica, Universidade Federal de Santa Catarina, CEP 88040-900, Florian\'{o}polis, SC, Brazil}
\author{S. P. Walborn}
\email{swalborn@udec.cl}
\affiliation{Departamento de F\'{i}sica, Universidad de Concepci\'on, 160-C Concepci\'on, Chile}
\affiliation{Millennium Institute for Research in Optics, Universidad de Concepci\'on, 160-C Concepci\'on, Chile}

\date{\today}

\begin{abstract}
We introduce the Twisted Gaussian Schell (TGS) state, a two-mode mixed Gaussian
state defined as the quantum-optical analog of the Twisted Gaussian Schell-model
beam of classical paraxial optics, characterized by the so-called twist phase. In the TGS state, the twist parameter arises when an asymmetric two-mode thermal state is subject to local
squeezing after the action of phase shifters and a beam splitter. Its defining quantum feature is
nonclassicality: although the state is separable in its natural bipartition, when the twist parameter is nonzero there are global quadratures that can be squeezed below the shot-noise limit. The nonclassicality has also  a direct signature in the joint photon-number distribution, which we obtain
in closed form. Moreover, coupling each mode to an ancillary vacuum at a balanced beam splitter yields a four-mode state with entanglement in select $2\times2$ bipartitions, with local description given by two TGS states, and all $1\times3$ bipartitions.
 For fixed input squeezing, increasing the twist parameter activates
entanglement where the state is otherwise separable and deepens it where already present. The classical physicality bound on the twist parameter
coincides with the quantum physicality condition. These results
advance the two-way bridge between classical beam engineering and quantum
information.
\end{abstract}

\maketitle

\section{Introduction}
A well-established formal analogy connects the transverse spatial structure of paraxial optical beams in classical optics with the state space of bosonic modes in quantum optics. Within the paraxial approximation, the slowly varying envelope of a monochromatic optical field obeys an equation formally equivalent to the Schr\"{o}dinger equation for a free particle, with the propagation coordinate playing the role of time. As a consequence, the transverse position and wave-vector degrees of freedom form a canonical pair obeying the same symplectic structure that underlies the quadrature operators of continuous-variable (CV) quantum systems \cite{Simon1987,Simon1988}.
\par
Within this correspondence, the Gaussian formalism central to continuous-variable (CV) quantum information theory~\cite{weedbrook2012,Serafini2017,Adesso2014} admits a direct classical counterpart. Gaussian beams correspond to Gaussian states, characterized by the first and second moments of the quadrature operators, and the linear paraxial systems that transform them, such as lenses, fractional Fourier transformers, and astigmatic mode converters, act as the same symplectic operations that beam splitters, phase shifters, and squeezers implement in quantum optics. Both preserve the canonical commutation relations and map Gaussian states onto Gaussian states.
This shared symplectic structure provides a unified framework for analyzing classical structured beams and multimode quantum states \cite{Arvind1995}. It has enabled the transfer of tools from quantum information theory to classical optics \cite{Spreeuw2001,souza07,Aiello2015,Forbes2019,Shen2022,Wang2024,Rodriguez-Lara2025}, including entanglement-inspired measures of non-separability between spatial degrees of freedom, as well as the reverse translation of classical beam-engineering techniques to quantum state engineering and quantum information processing \cite{gomes09a,tasca11,boucher2015}. 
\par
Such cross-fertilization has proven fruitful in spatial-mode encoding and in classical and quantum interferometry~\cite{malik2026highdimensionalquantumphotonicsroadmap}. Concepts from quantum entanglement theory, in particular, have been used to characterize correlations between orthogonal spatial modes of classical beams, yielding classically non-separable states that reproduce structural features of multipartite quantum systems~\cite{oliveira05,Borges2010,Chowdhury13,Balthazar2016}. Two-mode CV quantum states with vortex structure analogous to orbital angular momentum (OAM) beams of classical light have been studied extensively~\cite{agarwal97,Agarwal_2011,Puentes2021,Zhu2012a,Zhu2012b,Barral16,Li2015,Bandyopadhyay2011a,Bandyopadhyay2011b}, predominantly as pure, non-Gaussian states. Closer to the present work, two-mode Gaussian states carrying an angular-momentum (twist) coupling have also been considered, both as pure rotating wave packets~\cite{Dodonov_2015} and as the vacuum and thermal states of two harmonic modes coupled by angular momentum~\cite{rebon2011}. 
These studies have clarified both the power and the limits of the classical-quantum analogy and motivated the search for structured beams whose statistics admit a novel quantum-optical interpretation.
\par
Among structured partially coherent fields, Twisted Gaussian Schell-model (TGSM) beams \cite{simon93,simon98} occupy a particularly interesting position. The Gaussian Schell-model (GSM) describes partially coherent beams whose cross-spectral density is Gaussian in both spatial coordinates and coherence function. TGSM beams extend this model by incorporating a twist phase that arises from the coupling of orthogonal transverse coordinates in the form $x p_y - p_x y$, associated with OAM and rotational phase structure \cite{Friberg1994,Serna2001,ferreira2026secondordermomentequivalencetwisted}. 
\par
In classical optics, TGSM beams have attracted attention for their rich propagation dynamics and robustness~\cite{WangCai2010}. The interplay of partial coherence and phase structure governs beam spreading, rotation, and resilience to turbulence~\cite{Wang2012}, making them promising for optical manipulation~\cite{Zhao2009}, imaging through scattering media~\cite{Cai2009,Tong2012}, and free-space communication~\cite{WangCai2010,Wang2012}, while in nonlinear optics the coherence and twist influence phase matching in processes such as second-harmonic generation~\cite{deOliveira24} and parametric down-conversion~\cite{Hutter20,Hutter21,Santos2022}. Despite this body of work, a quantum-optical analog of TGSM fields as a bipartite quantum state has not been developed, and the properties of such a state remain largely unexplored~\cite{Hutter20,Ponomarenko2021,Hutter21}.
\par
In this work we exploit the formal equivalence between classical paraxial optics and continuous-variable quantum systems to introduce the Twisted Gaussian Schell (TGS) state, the quantum-state analog of the classical TGSM beam. We show that the state can be generated from an asymmetric two-mode thermal state by local squeezing between passive linear-optical networks.

The defining quantum feature of the TGS is two-mode nonclassicality. Although separable in its natural bipartition—as we confirm using the positive-partial-transpose (PPT) criterion~\cite{Simon2000}—it can exhibit two-mode squeezing below the shot-noise limit for nonzero twist. The twist controls the quadrature angle and can enhance the degree of squeezing. We investigate several properties of the TGS, including its nonclassicality, classical correlations, and joint photon-number distribution, all of which we obtain in closed form. We show that its nonclassicality can be converted into entanglement while preserving the twist structure by mixing each mode with an ancillary vacuum mode at a balanced beam splitter. Although the twist alone does not generate entanglement, it increases the entanglement in certain $2\times 2$ bipartitions and in every $1\times 3$ bipartition of the resulting four-mode state. Importantly, for the $2\times 2$ entangled bipartitions, the local description is given by TGS states that retain the structure of the input state at the beam splitter.
\section{Twisted Gaussian-Schell States}

In analogy with the Twisted Gaussian Schell model beams well-known in classical paraxial optics \cite{simon93,simon93b,simon98}, we define the Twisted Gaussian-Schell state $\hrho_{TGS}$, that is, a mixed Gaussian state whose Wigner function is (here we set $\hbar=1$):
\beq
W_{TGS}(\bx)=\frac{1}{(2\pi)^2\sqrt{\det\mV_{TGS}}}\exp\{-\frac{1}{2}\bx^\tp\mV_{TGS}\bx\}
\eeq
where we define the dimensionless phase space variables using mode order, 
\beq
\bx=(x_1,p_1,x_2,p_2)^\tp,
\eeq
with $\tp$ meaning transposition.
Here we will use a parametrization given in Ref.~\cite{wang19}, which is well suited for the quantum optical analog we pursue. The covariance matrix (CM) is:
\beq
\mVb_{TGS}= \begin{pmatrix}
     \Omega\kappa & 0 &0&v \\
      0 &   \Omega^{-1}\kappa &-v&0\\
     0 &-v& \Omega\kappa&0\\
    v  &0&0& \Omega^{-1}\kappa
   \end{pmatrix}.
   \label{eq:covmatVb}
\eeq
The CM~\eqref{eq:covmatVb} coincides with that of a TGSM beam derived in Ref.~\cite{wang19} (their Eq.~(4)), under the explicit identifications
\bse
\label{eq:dict}
\bea
\kappa &=& \omega_0^2\!\left(\frac{1}{k^2\omega_0^2}\!\left(\frac{1}{4\omega_0^2}+\frac{1}{\delta_0^2}\right)+\mu_0^2\right)^{1/2},\\
\Omega &=& \left(\frac{1}{k^2\omega_0^2}\!\left(\frac{1}{4\omega_0^2}+\frac{1}{\delta_0^2}\right)+\mu_0^2\right)^{-1/2},\\
v &=& \mu_0\omega_0^2,
\eea
\ese
where $\omega_0$ is the beam width, $\delta_0$ the transverse coherence length, $\mu_0$ the twist phase, and $k=2\pi/\lambda$ the wavenumber. The classical bound $|\mu_0|\leq (k\delta_0^2)^{-1}$ on the twist phase translates into the quantum physicality condition that we derive below, confirming that the two constraints are one and the same physical requirement expressed in different languages.
\par
In the context of quantum optics, the CM \eqref{eq:covmatVb}
describes a symmetric two--mode Gaussian state with identical local marginals and phase--sensitive inter--mode correlations.
The three parameters have clear physical interpretations:
$\kappa$ sets the overall mixedness of each local mode, $\Omega$ controls the amount of local squeezing, and the twist parameter $v$ determines the strength of inter--mode correlations, coupling $x_1$ with $p_2$ and $p_1$ with $x_2$.
The purity of the state is
\beq
\beta=\Tr(\hrho_{TGS}^2)=\left(\frac{1}{2}\right)^2\frac{1}{\sqrt{\det(\mV_{TGS})}}=
\frac{1}{4(\kappa^2-v^2)}.
\label{eq:purity}
\eeq
Since $\Tr(\hrho_{TGS}^2)
\leq 1$, 
the following conditions must be satisfied: 
\bse
\label{eq:condk}
\bea
\label{eq:cond1k}
\kappa+ v&\ge&\frac{1}{2},\\
\kappa- v&\ge&\frac{1}{2}
\label{eq:cond2k}.
\eea
\ese
For $\kappa=\tfrac12$ the local states are minimum--uncertainty (pure), while $\kappa>\tfrac12$ corresponds to added thermal noise. 
It is worth noting that the twist parameter $v$ is necessarily zero for any 
pure TGS state, in direct analogy with the classical result that the twist 
phase vanishes in the fully coherent limit~\cite{simon93}.
To see this, recall that purity requires $\kappa^2 - v^2 = 1/4$, 
i.e.\ $\kappa = \sqrt{v^2 + 1/4}$, while physicality requires 
$\kappa - |v| \geq 1/2$.
Substituting the purity condition into the physicality constraint leads to $|v| \leq 0$, and hence $v = 0$. 
The unique pure TGS state is therefore the two-mode vacuum 
($v=0$, $\kappa = 1/2$).
Thus, the quantum twist parameter 
$v$ is bounded by the excess noise $\kappa - 1/2$ above the vacuum 
and vanishes for pure states.
Any nonzero twist therefore necessarily implies a mixed global state, 
confirming that the TGS state is an intrinsically mixed-state object 
with no pure-state analogue beyond the vacuum.

\subsection{Generating TGS states}
\par
The Williamson form (WF) of $\mV_{TGS}$~\cite{williamson}, with $\kappa\ge v$ required for positive definiteness, is: 

\beq
\mVb_{th}= \begin{pmatrix}
    \kappa-v & 0 &0&0 \\
      0 &  \kappa-v  &0&0\\
     0 &0&  \kappa+v &0\\
    0 &0&0&  \kappa+v 
   \end{pmatrix}=\mSb\mVb_{TGS}\mSb^\tp.
\eeq
Here
$\mS$ is the symplectic matrix:

\beq
\mSb=\begin{pmatrix}
    0 & \sqrt{\tfrac{\Omega}{2}} &\frac{1}{\sqrt{2\Omega}}&0 \\
      -\frac{1}{\sqrt{2\Omega}}&0   &0&\sqrt{\tfrac{\Omega}{2}}\\
  0&   - \sqrt{\tfrac{\Omega}{2}} &\frac{1}{\sqrt{2\Omega}} &0\\
     \frac{1}{\sqrt{2\Omega}}  &0&0&  \sqrt{\tfrac{\Omega}{2}}
   \end{pmatrix}.
   \label{eq:defSmb}
\eeq
It is worth noting that $\mV_{th}$ is the covariance matrix of a two-mode thermal state $\hrho_{th}$.
There is thus a metaplectic evolution 
\beq
\hrho_{TGS}=\hUS\hrho_{th}\hUS^\dagger,
\label{eq:rhoTGS}
\eeq 
given by a unitary operator $\hUS$ (metaplectic operator) associated with the matrix $\mS$.
Note that the purity of $\hrho_{TGS}$, given in~\eqref{eq:purity}, is preserved by $\hUS$, and that 
$\hrho_{th}$ must also satisfy Eqs.~\eqref{eq:condk}. 
\par
\begin{figure}
   \centering
   \includegraphics[width=8cm,angle=0]{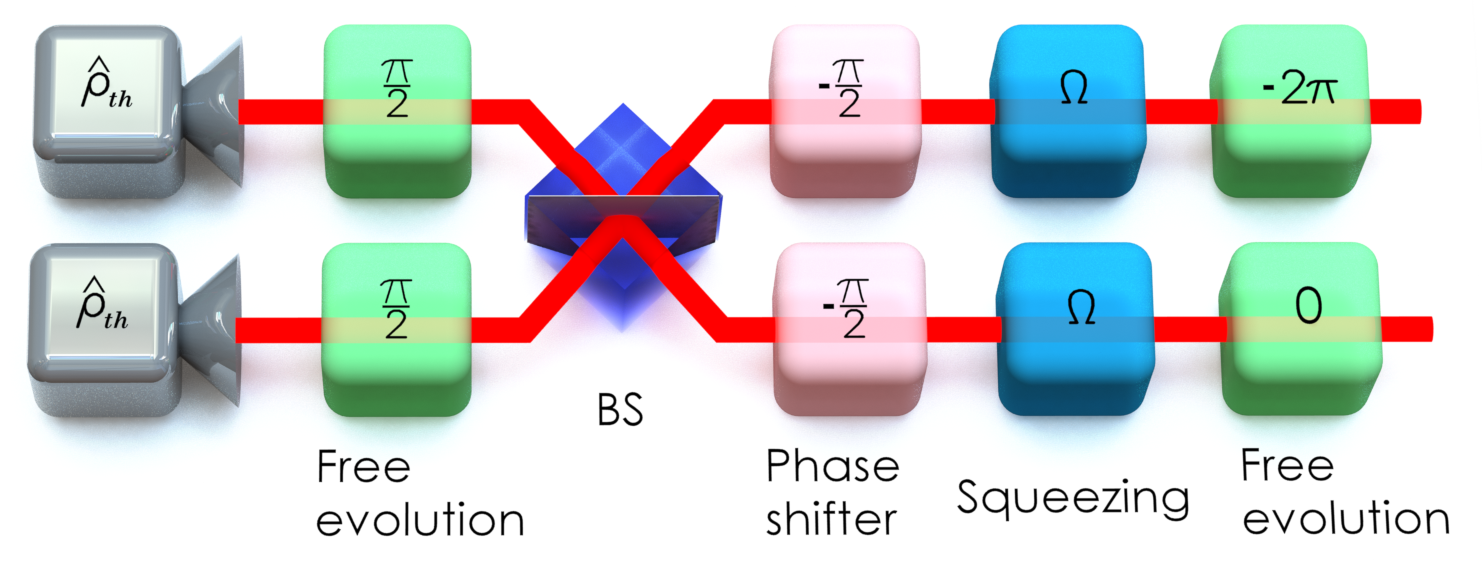}
   \caption{Scheme to generate the TGS state from two independent thermal modes.
From Eq.~\eqref{eq:defSmb}, the state is generated by the inverse
transformation $\mSb^{-1}=\mOb^{\prime -1}\mSb_d^{-1}\mOb^{-1}$ acting on the thermal
input. See text for details.}
   \label{Fig:gen}
\end{figure}

 
Eq.~\eqref{eq:rhoTGS} gives a recipe to construct the TGS state from an initial thermal state. 
In order to experimentally implement the coherent evolution $\hUS$, one must first find the Euler decomposition of the matrix 
$\mSb$, \ie 
\beq
\mSb=\mOb\mSb_d\mObp,
\label{eq:euler}
\eeq
where $\mOb, \mObp$ are orthogonal symplectic matrices, and we define the diagonal symplectic matrix 
\beq
\mSb_d=\diag(s_1,s_1^{-1},s_2,s_2^{-1}).
\eeq
Therefore we will have that 
\beq
\hUS=\hUO\hUSd\hUOp,
\eeq
where $\hUO$ and $\hUOp$ are metaplectic evolutions that can be implemented with beam splitters (BS), phase shifters (PS), and 
free evolution (FE), while $\hUSd$ is associated with local squeezing. In Appendix~\ref{AppEulerDecomp} we show that:
\bea
\label{eq:Sdec}
\mSb=& \mS_{FE}(-\frac{\pi}{2},-\frac{\pi}{2})\mS_{BS}(\frac{\pi}{2})\mS_{PS}(\frac{\pi}{2}) \times \\ \nonumber
& \diag\left(\frac{1}{\sqrt{\Omega}},\sqrt{\Omega},\frac{1}{\sqrt{\Omega}},\sqrt{\Omega}\right)
\mS_{FE}(2\pi,0).
\eea
A sketch of the TGS generation scheme from the thermal state $\rho_{th}$ is shown in Fig.~\ref{Fig:gen}. The Euler decomposition~\eqref{eq:Sdec} illuminates the physical origin of
the TGS correlations. Reading the generation scheme, the beam splitter acts first
on two thermal modes of unequal occupation (Williamson variances $\kappa-v$ and
$\kappa+v$), producing classical inter-mode correlations whose strength is set by
the imbalance $2v$ and which therefore vanish at $v=0$. The subsequent phase shifter, 
squeezing and free evolution are local. Each individual mode is squeezed below the shot-noise level only when
$\Omega < \frac{1}{2\kappa}$ or $\Omega > 2\kappa$, while inter-mode correlations depend entirely on the parameter $v$. The inter-mode correlations are thus classical in
origin, while any nonclassicality is supplied by the local squeezing acting on the
already-correlated modes. We further explore these points in the next section. 

Finally, we note that the generation scheme of Fig.~\ref{Fig:gen} is the CV quantum analog of the classical optical procedure in which an astigmatic GSM beam is transformed by a cylindrical-lens array into a TGSM beam~\cite{Friberg1994,wang19}. In both cases, the twist correlations are imprinted by a symplectic transformation of the same structural form, such that only its physical realization differs, cylindrical lenses in the classical setting versus squeezing and beam splitters in the quantum case.

\section{Separability, nonclassicality, and classical correlations}
\label{sec:separabillity}
  In this section we show that the TGS state is separable yet nonclassical, with the twist responsible for classical correlations that allow a two-mode quadrature to be squeezed below the shot-noise limit. We then determine the joint photon number distribution and evaluate the classical correlations in the Fock basis.
\subsection{Separability and classical correlations}
\label{sec:separability}

We first ask whether the TGS state is entangled in its natural two-mode bipartition. Because $\hrho_{TGS}$ is a
Gaussian state, we may apply the PPT criterion for continuous-variable
systems~\cite{Simon2000}. 
 
The partial transposition with respect to a bipartition consists into flip the momenta of one of the parties, implemented by a diagonal matrix $\Lambda$. Thus, the covariance matrix associated with the partial transpose density operator is $\tilde{\mVb}=\Lambda\mVb\Lambda$, and the original state is entangled if the minimum symplectic eigenvalue $\tilde\nu_{min}$ of
$\tilde{\mVb}$ satisfies $\tilde\nu_{min} < \tfrac12$. The symplectic eigenvalues of a $2n\times2n$ covariance matrix
$\tilde{\mVb}$ are the moduli of the eigenvalues of $\imath \tilde{\mVb} \mJb$ ~\cite{Serafini2017,Adesso2014}, with
\begin{align}
\mJb=\oplus_{j=1}^n \mJb_j\quad;\quad\mJb_j=\begin{pmatrix} 0& 1\\ -1& 0\end{pmatrix}.
\label{eq:J}
\end{align}
However, for the TGS state it suffices to invoke the lemma of Ref.~\cite{Simon2000}:
a two-mode Gaussian state with block-structured covariance matrix
\beq
\mVb=   \begin{pmatrix} A & C \\
C^T & B \\
\end{pmatrix}
\eeq
is separable iff $\mathrm{det}\,C \geq 0$ . Inspecting the TGS covariance
matrix~\eqref{eq:covmatVb} gives $\mathrm{det}\,C = v^2 \geq 0$ for all allowed
parameter values, so $\hrho_{TGS}$ is separable for every $v$. The 
inter-mode correlations induced by $v$ are therefore correlations of a separable
mixed state. They can be quantified using the 
mutual information $I(\hat{\rho}_{AB})=S(\hat{\rho}_A)+S(\hat{\rho}_B)-S(\hat{\rho}_{AB})$, where $\hat{\rho}_A$ and $\hat{\rho}_B$ are the marginal quantum states of the bipartite state $\hat{\rho}_{AB}$, and $S(\hat{\rho})=-\Tr(\hat{\rho}\ln(\hat{\rho}))$ is the von Neumann entropy. Since the TGS state is Gaussian, its mutual information is computed as~\cite{Serafini2017,Adesso2014}: 
\begin{align}
I_G=2\,h(\kappa)-h(\kappa-v)-h(\kappa+v),
\label{eq:IG}
\end{align}
with $h(\nu)=(\nu+\tfrac12)\log_2(\nu+\tfrac12)-(\nu-\tfrac12)\log_2(\nu-\tfrac12)$ the
bosonic entropy in bits. $I_G$ is independent of the squeezing $\Omega$, vanishes at $v=0$ and increases monotonically with $v$.
\subsection{Nonclassicality}
\label{sec:nonclassicality}
Separability does not, however, imply classicality. For a Gaussian state the
relevant notion of classicality is the positivity of the Glauber--Sudarshan $P$
function \cite{vogel2000}, which holds if and only if all quadrature variances are at least as large as the vacuum variance: $\mVb_{TGS}\geq\tfrac12\mathbb{1}$. The eigenvalues of the
covariance matrix~\eqref{eq:covmatVb} are doubly degenerate and given by
\begin{equation}
\lambda_{\pm}=\frac{1}{2} \left[ \kappa F
\pm\sqrt{\kappa^2(F^2-4)+4v^{2}} \right],
\label{eq:lammin}
\end{equation}
where we define 
\begin{equation}
\label{eq:defF}
F=\Omega+\frac{1 }{\Omega}
\end{equation}
as a global squeezing parameter.
It is straightforward to check that $\lambda_{\min}=\lambda_{-} \leq \lambda_{+}$ for all $\Omega>0$ with $\Omega\neq 1$, $\kappa\ge \frac{1}{2}+v$ and $v> 0$, with the equality $\lambda_{\min}=\lambda_{-}= \lambda_{+}$ holding for $\Omega=1$, \ie $F=2$, and $v=0$. 
\par
Now, note that $\lambda_{-}=\lambda_{-}(F)\geq 0$, with equality attained in the limit $\lim_{F\rightarrow +\infty}\lambda_{-}(F)=0$. Also,
we have that $\pdv{\lambda_{-}}{F}=\frac{1}{2} \left(\kappa -\frac{F \kappa ^2}{\sqrt{\left(F^2-4\right) \kappa ^2+4 v^2}}\right)< 0$
for all values of $F> 2$, $\kappa \geq \frac{1}{2}+v$ and $v > 0$. Thus, $\lambda_{-}(F)$ is strictly decreasing in $F$ on the interval $(2,\infty)$, attaining its maximum at 
$\lambda_{-}(F=2)=\kappa-v\geq \frac{1}{2}$ (see Eq.~\eqref{eq:cond2k}). The value $\lambda_{-}(F^*)=\frac{1}{2}$ is reached at
$F^*=\frac{1}{2\kappa}(1+4(\kappa^2-v^2))$.
Therefore, we have $\lambda_{-}<\frac{1}{2}$ if and only if 
\begin{align}
\label{eq:FlessFstar}
F> F^*=\frac{1}{2\kappa}\left(1+\frac{1}{\beta}\right),
\end{align}
with $\beta$ the purity defined in~\eqref{eq:purity}.
\par
The eigenvectors associated with the doubly degenerate eigenvalue $\lambda_{\min}=\lambda_{-}$ of $\mVb_{TGS}$ are
\bse
\label{eq:globalquadra}
\begin{align}
{\bf v}_{1}&=
\left\{
\begin{matrix}
\left(\frac{\lambda_{-}}{v}-\frac{\kappa}{v\Omega},0,0,1\right)^\tp& \Omega\geq 1\\
\left(1,0,0,\frac{\lambda_{-}}{v}-\frac{\kappa\Omega}{v}\right)^\tp& 0<\Omega\leq 1
\end{matrix}\right.,\\
{\bf v}_{2}&=
\left\{
\begin{matrix}
\left(0,1,-\left(\frac{\lambda_{-}}{v}-\frac{\kappa}{v\Omega}\right),0\right)^\tp& \Omega\geq 1\\
\left(0,-\left(\frac{\lambda_{-}}{v}-\frac{\kappa\Omega}{v}\right),1,0\right)^\tp& 0<\Omega\leq 1
\end{matrix}\right..
\end{align}
\ese
These constitute global quadratures for all values of $F>2$, $\kappa\geq \frac{1}{2}+v$ and $v>0$. Hence, $F$ acts as a kind of global squeezing parameter such that, whenever it satisfies Eq.~\eqref{eq:FlessFstar}, for $v\neq 0$, it signals the presence of squeezing, \ie $\lambda_{\min}<\frac{1}{2}$, in the global quadratures in Eqs.~\eqref{eq:globalquadra}.
\par
For small values of the twist parameter $v$, we have
\begin{align}
\lambda_{min}=\lambda_{-}=\left\{
\begin{matrix}
 \frac{\kappa }{\Omega } +O(v^2) & \Omega \geq 1 \\
 \kappa  \Omega +O(v^2) & 0<\Omega\leq 1 \\
\end{matrix}
\right.,
\end{align}
and therefore the eigenvectors in Eq.~\eqref{eq:globalquadra} become
\bse
\begin{align}
{\bf v}_{1}&=\left\{
\begin{matrix}
\left(O(v^2),0,0,1\right)^\tp& \Omega \geq 1 \\
\left(1,0,0,O(v^2)\right)^\tp
& 0<\Omega\leq 1 \\
\end{matrix} \right.\;\;, \\
{\bf v}_{2}&=\left\{
\begin{matrix}
\left(0,1,O(v^2),0\right)^\tp& \Omega \geq 1 \\
\left(0,O(v^2),1,0\right)^\tp
& 0<\Omega\leq 1 \\
\end{matrix}\right.\;\;.
\end{align}
\ese
\noindent
Thus, when the twist parameter $v$ vanishes, there is only local squeezing: at ${\bf v}_{1}=(0,0,0,1)^\tp$ and ${\bf v}_2=(0,1,0,0)^\tp$ when $\Omega>1$, in which case $\lambda_{\min}=\frac{\kappa}{\Omega}<\frac{1}{2}$; or at ${\bf v}_{1}=(1,0,0,0)^\tp$ and ${\bf v}_2=(0,0,0,1)^\tp$ when $0<\Omega<1$, in which case $\lambda_{\min}=\kappa\Omega<\frac{1}{2}$.
\par

The quantity $\lambda_{\min}$ is the key parameter governing most results that follow.
The twist acts on it as a resource, since $\partial\lambda_{\min}/\partial v<0$,
so at fixed thermal noise $\kappa$ and global squeezing $F$, increasing
$v$ lowers $\lambda_{\min}$ monotonically and, for $\Omega\neq1$, can drive it below
$\tfrac12$. Squeezing ($\Omega\neq1$) is required for any nonclassicality, but at fixed squeezing the twist is the trigger that carries $\lambda_{\min}$ across the $\tfrac12$ threshold. The same crossing turns on sub-shot-noise squeezing here and, after the four-mode embedding of Sec.~\ref{sec:entanglement}, entanglement, because both are monotone in $\lambda_{\min}$. See section \ref{sec:entanglement} and Fig. \ref{fig:kF-Tplots} for further discussion on nonclassicality/entanglement.
\par
\subsection{Two-mode entangling scheme}
Nonclassicality and entanglement thus part ways in the TGS state, which exhibits
two-mode squeezing without entanglement.

Nonclassicality is nonetheless a resource for entanglement, since any nonclassical
Gaussian state can be entangled by passive linear optics alone~\cite{Kim2002,Asboth2005,Wolf2003}.
For a two-mode input, the maximum entanglement, measured by the logarithmic negativity, attainable in this way
is~\cite{Wolf2003}
\begin{equation}
E_\mathcal{N}^{\max}=\max\!\left(0,\,-\log_2 2 \sqrt{\lambda_1\lambda_2}\right),
\label{eq:WEPbound}
\end{equation}
with $\lambda_1,\lambda_2$ the two smallest eigenvalues of the covariance matrix.
 This bound is saturated by applying the following passive metaplectic operations on the TGS state: a
$\tfrac{\pi}{2}$ phase shift on one mode, characterized by the symplectic matrix $\mJb_1\oplus \mathbb{1}_2$, followed by a balanced 50/50 beam splitter that
mixes the two TGS modes. The output covariance matrix $\mVb^\prime=\mS_{BS}(\frac{\pi}{2})\mJb_1\oplus \mathbb{1}_2\mVb_{TGS}(\mJb_1\oplus \mathbb{1}_2)^\tp\mS^\tp_{BS}(\frac{\pi}{2})$ is in Simon normal form~\cite{Simon2000}, \ie
\begin{equation}
\mVb^\prime=\begin{pmatrix} a'\,\mathbb{1}_2 & \diag(d,-d)\\[2pt]
\diag(d,-d) & b'\,\mathbb{1}_2\end{pmatrix},
\label{eq:Vprime}
\end{equation}
with $a'=\tfrac{\kappa F}{2}{+}v$, $b'=\tfrac{\kappa F}{2}{-}v$, and $d=\tfrac{\kappa}{2}(\Omega-\Omega^{-1})$. 
The partial transpose of $\mVb^\prime$ has symplectic eigenvalues $\{\lambda_{-},\lambda_{+}\}$ with $\lambda_{\pm}$ in Eq.~\eqref{eq:lammin}, the smallest being $\lambda_{-}$, and the logarithmic negativity in Eq.~\eqref{eq:WEPbound} becomes,
\begin{equation}
E_\mathcal{N}^{\max}=\max\!\left(0,\,-\log_2 2\lambda_{\min}\right),
\label{eq:EN2mode}
\end{equation}
where we set $\lambda_{-}=\lambda_{min}$ following Section~\ref{sec:nonclassicality}. 
\par
This scheme is optimal for entanglement, but it does not preserve the structure of the TGS. The twist has been rotated into the local variances $a'$ and $b'$, while the correlation block $\diag(d,-d)$ depends only on $\kappa$ and $\Omega$, corresponding to a two-mode squeezed state in which the Twisted Gaussian Schell structure no longer appears. Because our aim is the quantum-optical analog of a partially coherent structured light beam, in Sec.~\ref{sec:entanglement} we present a four-mode realization that reaches the same optimal $E_\mathcal{N}$ while keeping the twist intact, and analyze the bipartition structure in detail.

\subsection{Joint photon number distribution}
\label{sec:fockbasis}
Having established that the TGS state is separable yet nonclassical, we now examine its photon-number distribution $P(n_1,n_2)$, corresponding to the outcome statistics of number-resolving detection on the two modes, a directly accessible characterization of the state. In what follows we show that the twist parameter renders the photon-number distribution non-factorable, $P(n_1,n_2)\neq P(n_1)P(n_2)$ whenever $v\neq 0$, and we obtain $P(n_1,n_2)$ in closed form. We quantify the Fock-basis correlations via the mutual information.
\begin{figure} 
    \centering
    \includegraphics[width=8.8cm]{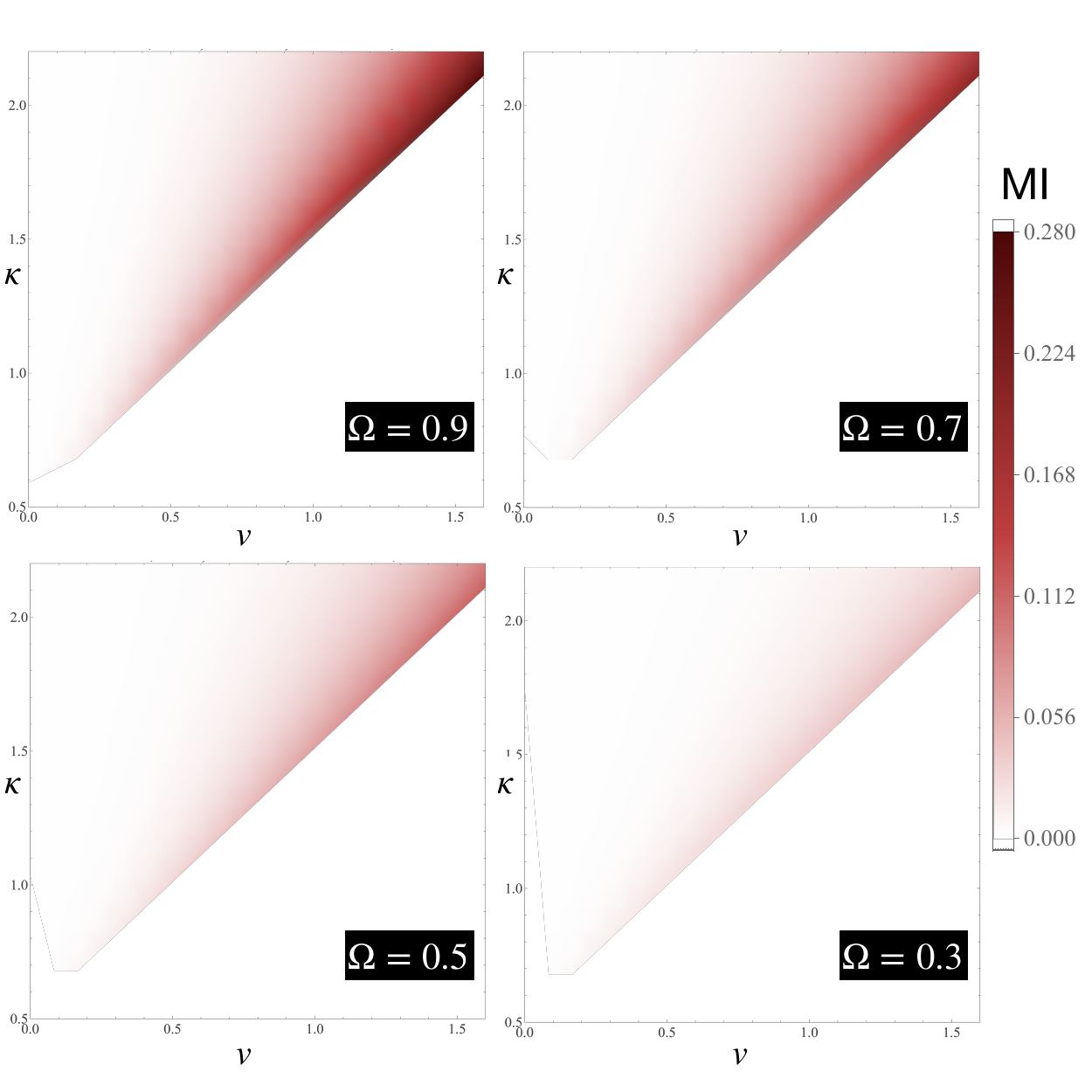}
    \caption{Density plots of the mutual information in Eq.~\eqref{eq:I} for the joint photon distribution $P(n_1,n_2)$ in Eq.~\eqref{eq:finalPn1n2}, and marginals in~\eqref{eq:formulaforPnj}, of a TGS state. See text for explanation.}
    \label{fig:FigHMMI}
\end{figure}
\par 
The joint probability for detection in the Fock basis is
\bea
P(n_1, n_2)&=&\matrixel{\no\nt}{\hrho_{TGS}}{\no\nt} \\ \nonumber
&=&
\int d\bx\;W_{TGS}(\bx)
\matrixel{\no\nt}{\hR_{\bx}}{\no\nt}\nonumber\\
&=&\int d\bxo\int d\bxt \;W_{TGS}(\bxo,\bxt)\times \\ \nonumber 
&\times & \matrixel{\no}{\hR_{\bxo}}{\no}\otimes \matrixel{\nt}{\hR_{\bxt}}{\nt}.
\label{eq:Pn1n2Laguerre}
\eea
where $\frac{1}{2^2}\hR_{\bx_j}$, $j=1,2$, is the reflection operator in mode $j$.
The matrix elements $\matrixel{n_j}{\hR_{\bx_j}}{m_j}$ were calculated in~\cite{Groenewold1946} and for 
the diagonal elements we have:
\beq
\matrixel{n_j}{\hR_{\bx_j}}{n_j}=2\,(-1)^{n_j}\,e^{-(x_j^2+p^2_j)}\;L^0_{n_j}(2(x_j^2+p^2_j)),
\eeq
 where $L^0_{n_j}=L_{n_j}$ are the simple Laguerre polynomials. 
Using this and the defintion of the TGS Wigner function, we have 

\bea
P(n_1, n_2)&=&\frac{(-1)^{n_1+n_2}}{(\pi)^2}C \iiiint dx_1 dp_1 dx_2 dp_2 \nonumber\\
&
\times & e^{-a(x_1^2+x_2^2)-b(p_1^2+p_2^2)+c(x_1p_2-p_1x_2)}\nonumber\\
&\times & L_{n_1}(2(x_1^2+p_1^2))L_{n_2}(2(x_2^2+p_2^2)),
\label{eq:Pn1n2ini}
\eea
where we define
\bse
\label{eq:defabc}
\bea
a&=&\frac{\kappa}{2\Omega(\kappa^2-v^2)} +1 ,\\
b&=&\frac{\kappa\Omega}{2(\kappa^2-v^2)}+1 ,
\label{eq:defb}\\
c&=&\frac{v}{\kappa^2-v^2},
\label{eq:defc}\\
C&=&\frac{1}{\kappa^2-v^2}.
\label{eq:defC}
\eea
\ese
It is immediately apparent that conditions \eqref{eq:cond1k}, \eqref{eq:cond2k} and $\Omega>0$ lead to:
\bse
\label{eq:conabc}
\begin{align}
1&<a,b\leq \infty,
\label{eq:cona}\\
|c|&<1.
\end{align}
\ese 
\par
In Appendix~\ref{App:derPn1n2}, we show that
\begin{align}
P(n_1,n_2) &= \frac{C}{a\,\eta\, 2^{2(n_1+n_2)}}
\left(\frac{2-\eta}{\eta}\right)^{n_1+n_2} 
\sum_{k_1=0}^{n_1} \sum_{k_2=0}^{n_2}\nonumber\\
&\times B^{k_1+k_2} \,
\mathcal{F}_{n_1}(k_1) \,
\mathcal{F}_{n_2}(k_2) \nonumber\\
&\times \mathcal{S}_{n_1}(k_1,k_2) \,
\mathcal{S}_{n_2}(k_2,k_1),
\label{eq:finalPn1n2}
\end{align}
with
\bse
\begin{align}
\mathcal{F}_n(k) &= 4^n \frac{\Gamma(n-k+\tfrac{1}{2})\,\Gamma(k+\tfrac{1}{2})}{\pi}, \\[4pt]
\mathcal{S}_n(k,x) &= \frac{1}{(n-k)!\,x!} \,
{}_{2}F_{1}\!\left(-(n-k), -x; \frac{1}{2}; \frac{A}{4}\right),
\end{align}
\ese
where $\Gamma$ is the gamma function, $_2F_1(-p,-q;\frac{1}{2};\frac{A}{4})=\sum_{j=0}^{\min(p,q)}\frac{(-p)_j(-q)_j}{\left(\frac{1}{2}\right)_jj!}\left(\frac{A}{4}\right)^j$ is the Gauss hypergeometric polynomial with $p$ and $q$ non-negative integers, and we define
\begin{align}
\eta&=b-\frac{c^2}{4a},
\label{eq:defeta}\\
 B&=\frac{4ab-8b-c^2}{4ab-8a-c^2 },\\
 A&=B\frac{64 c^2 }{(4ab-8b-c^2)^2}.
\end{align}
The conditions in Eqs.~\eqref{eq:conabc} imposes that $1<\eta<\infty$ while $-\infty\leq A,B\leq \infty$. 
\par
From Eq.~\eqref{eq:Pn1n2ini} we immediately realize that 
the change $x_1\leftrightarrow p_1$ and $x_2\leftrightarrow p_2$
is equivalently to $\Omega\rightarrow \frac{1}{\Omega}$ and $v\rightarrow -v$ or $a\leftrightarrow b$ and $c\rightarrow -c$.
This means that $P(n_1,n_2)=P_{\Omega,v}(n_1,n_2)=P_{1/\Omega,-v}(n_1,n_2)$
or equivalently $P(n_1,n_2)=P_{a,b,c}(n_1,n_2)=P_{b,a,-c}(n_1,n_2)$.
Therefore, it is enough to analyze $P(n_1,n_2)$ in the regime of parameters: 
\bse
\label{eq:constbetaalv}
\begin{align}
&0<\Omega \leq 1,
\label{eq:constbetaalv1}\\
&v\geq0,\\
&\kappa\geq \frac{1}{2}+|v|.
\end{align}
\ese
Note that $\Omega=1$ corresponds to a TGS state without squeezing while in the limit $ \Omega\rightarrow 0$ the squeezing is infinite.
\par
The Fock-basis correlations can be understood physically as a direct consequence of the cross-quadrature coupling $c(x_1 p_2 - p_1 x_2)$ in the exponent of Eq.~\eqref{eq:Pn1n2ini}.
When $v=0$ we have from~\eqref{eq:defc} that $c=0$, so it is immediately evident from~\eqref{eq:Pn1n2ini} that the joint probability factorizes $P(n_1, n_2)=P(n_1)P(n_2)$, so there are no correlations. When $v\neq 0$ the cross-quadrature coupling prevents the integrand in~\eqref{eq:Pn1n2ini} from factorizing, so the joint distribution $P(n_1,n_2)$ cannot be written as a product of marginals.
\begin{figure}[t] 
   \centering
   \includegraphics[width=9cm]{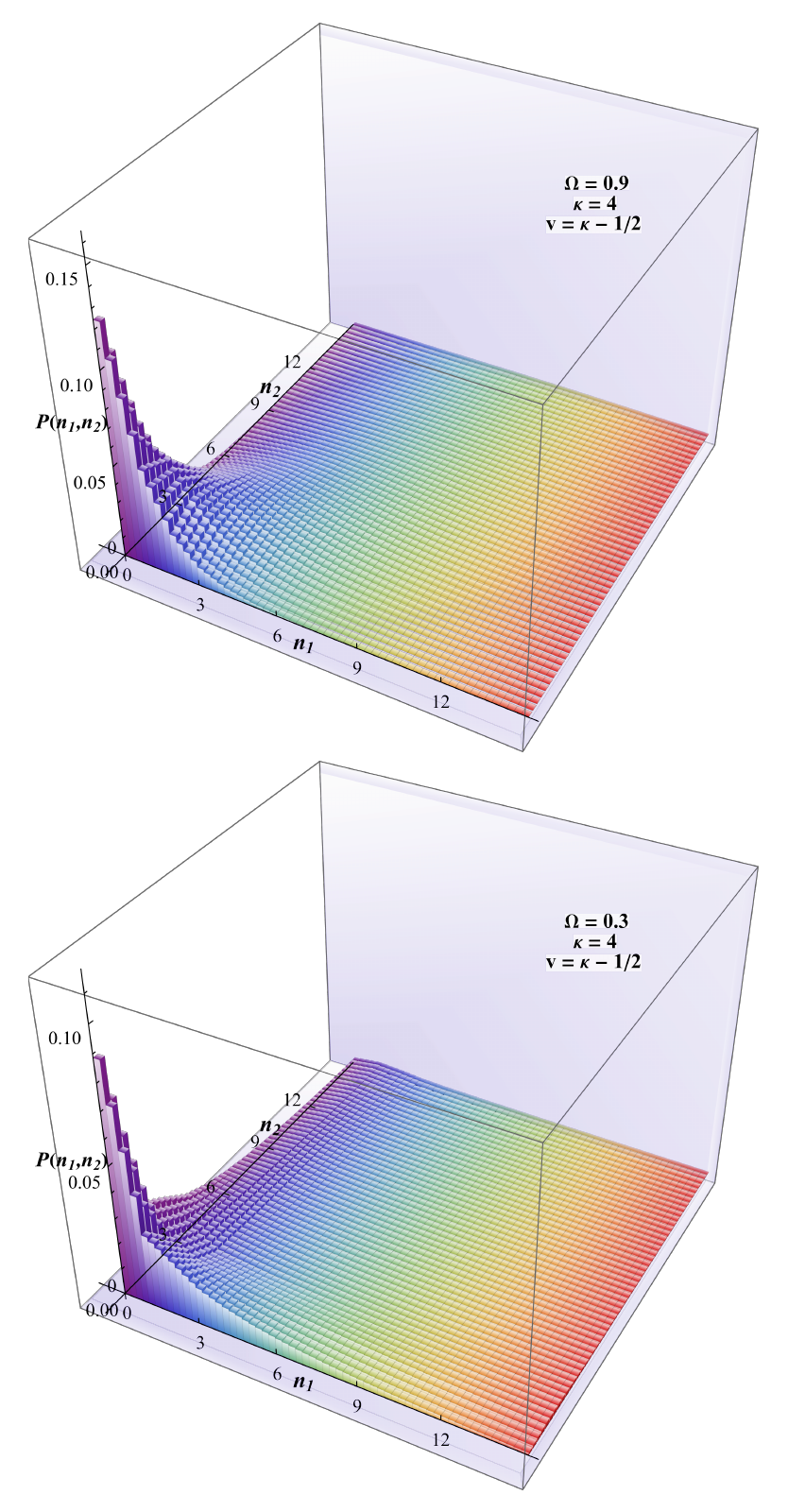} 
   \caption{Joint photon number distribution $P(n_1,n_2)$ of a TGS state at maximum correlations $v=\kappa-\frac{1}{2}$ with $\kappa=4$. On top for small squeezing , corresponding to $\Omega=0.9$, and at the bottom for large squeezing, corresponding to $\Omega=0.5$.}
   \label{fig:pn1n2plots}
\end{figure}
\par 
The correlations present in the joint number distribution can be quantified by the mutual information, given by:
\beq
I=\sum_{n_1=0}^{+\infty}\sum_{n_2=0}^{+\infty}\;P(n_1,n_2)\;\log_2\left(\frac{P(n_1,n_2)}{P_{n_1}(n_1)P_{n_2}(n_2)}\right),
\label{eq:I}
\eeq
where the marginal probabilities distributions are (see Appendix~\ref{App:derPnj} for the derivation):
\begin{align}
P(n_j) &= D \,
(uw)^{\frac{n_j}{2}} \, P_{n_j}\!\left( \frac{u+w}{2\sqrt{uw}} \right),
\label{eq:formulaforPnj}
\end{align}
with $j=1,2$, $P_n$ is the Legendre polynomial~\cite{Lebedev1965}, and we define the parameters:
\bse
\begin{align}
D&=\frac{2(4 (a-1) (b-1)- c^2)}
{\sqrt{\left(4 ab-4 a-c^2\right)\left(4 ab-4 b-c^2\right)}},\\
u&= \frac{8(b-1)}{4ab - 4a - c^2} - 1,
\label{eq:defu}\\
w&=\frac{8(a-1)}{4ab - 4b - c^2} - 1.
\label{eq:defw}
\end{align}
\ese
Eq.~\eqref{eq:formulaforPnj} corresponds to the photon number distribution of a one-mode squeezed thermal state \cite{dodonov1994}.
\par
In Figure~\ref{fig:FigHMMI}, we show the behavior of the Fock-basis mutual information 
as a function of $\kappa$, $v$, and $\Omega$. For a fixed value of the 
squeezing parameter $\Omega$, the mutual information attains its maximum 
at $v=\kappa-\frac{1}{2}$. 
As can be observed from the grayscale on the right of the plots in Figure~\ref{fig:FigHMMI}, for any fixed value of $\kappa$ and $v$, the mutual information is greater for weaker squeezing (larger $\Omega$). Across the four squeezing values shown in Fig.~\ref{fig:FigHMMI}, ranging from $\Omega=0.3$ to $\Omega=0.9$, this trend is monotonic: $I$ grows as $\Omega$ approaches $1$.
\par
The behavior of the joint photon number distribution 
when the correlations are maximal, i.e., $v=\kappa-\frac{1}{2}$, is shown 
in Figure~\ref{fig:pn1n2plots} for weak ($\Omega=0.9$) and strong squeezing ($\Omega=0.5$). For $\Omega=0.5$, 
the probabilities are larger for $n_1=0$ (for a fixed $n_2$) or $n_2=0$ (for a fixed $n_1$), whereas for 
$\Omega=0.9$ they are larger when $n_1\sim n_2$. 
Since the behavior of $P(n_1,n_2)$ around $n_1\sim n_2\sim 0$ is 
similar in both cases, the additional contribution from 
$n_1\sim n_2\neq 0$ increases the mutual information for 
$\Omega=0.9$. Thus, increased squeezing leads to a decrease 
in the classical Fock-basis correlations.
\par
It is illuminating to contrast these basis-dependent
correlations with the total quantum correlation quantified by the
 mutual information $I_G$ in~\eqref{eq:IG}. The mutual
information~\eqref{eq:I} corresponds to a particular (photon-number) measurement on the state, so
$I\le I_G$ by the data-processing inequality \cite{NielsenChuang2010}. Squeezing thus leaves the total quantum correlation $I_G$
unchanged but redistributes it among observables: as $\Omega$ moves away from $1$,
the photon-number basis captures a smaller share of the correlations, decreasing $I$ even
as the two-mode squeezing of Sec.~\ref{sec:separability} deepens. The
classical Fock-basis correlations are thus largest near $\Omega=1$, where the state
is classical ($\lambda_{\min}\ge\tfrac12$).
Nonclassicality and classical Fock-basis correlations therefore vary along
distinct axes, controlled by $\Omega$ and by $(\kappa,v)$, respectively.

\section{Twist-assisted entanglement in TGS state}
\label{sec:entanglement}
\begin{figure}
   \centering
   \includegraphics[width=8cm,angle=0]{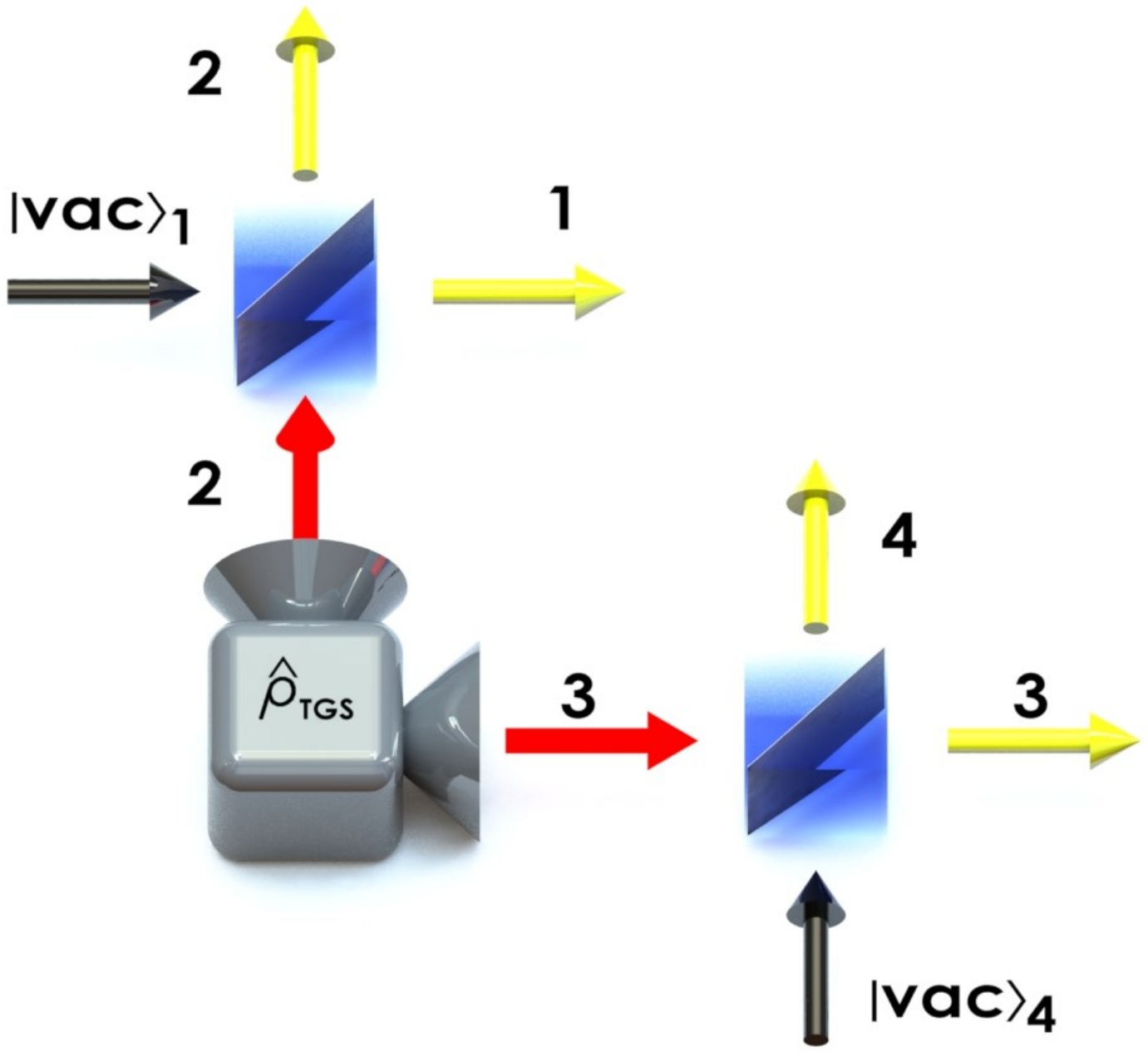}
   \caption{Scheme to activate entanglement from the TGS state. The two-mode TGS state (modes 2 and 3) is mixed with two ancillary vacuum modes (modes 1 and 4) at balanced beam splitters. The output four-mode state $\mVb_f$ is separable in the $12|34$ bipartition (since the beam splitters are local with respect to that cut) but can be entangled in the cross bipartitions $14|23$ and $13|24$, depending on the values of $\kappa$, $\Omega$, and $v$.}
   \label{Fig:entangled-state}
\end{figure}
The minimal scheme of Sec.~\ref{sec:nonclassicality} shows how recombining the two TGS modes on a BS can entangle them
optimally, but this eliminates the twist structure.
We now present a four-mode realization that reaches the same logarithmic
negativity without recombining the modes: each TGS mode is coupled only to its own
ancillary vacuum at a balanced BS.
\par
 We consider that the first beam splitter acts on modes $1$ and $2$ and the second beam splitter acts on modes $3$ and $4$, as shown in Fig.~\ref{Fig:entangled-state}. The {symplectic} matrix {corresponding} to the two beam splitters {in mode order} is:
    \begin{equation}
\mSb_{BS}(\theta_1,\theta_2)= \mSb_1(\theta_1) \oplus \mSb_2(\theta_2), 
\end{equation}
with
\bea
    \mSb_j(\theta_j) = 
    \begin{pmatrix}
  \cos\frac{\theta_j}{2}&0&\sin\frac{\theta_j}{2}&0
  \\
     0&  \cos\frac{\theta_j}{2}&0&\sin\frac{\theta_j}{2}\\
    -\sin\frac{\theta_j}{2} &0&  \cos\frac{\theta_j}{2}&0\\
    0 &-\sin\frac{\theta_j}{2}&0&  \cos\frac{\theta_j}{2}
     
   \end{pmatrix}.
\eea
\par
The covariance matrix of the initial state corresponding to vacuum states in modes $1$ and $4$ and a TGS state in modes $2$ and $3$ is:
\bea
\mVb_{i}=
   \begin{pmatrix}
   \frac{1}{2}&0&0&0&0&0&0&0\\
     0&  \frac{1}{2}&0&0&0&0&0&0\\
     0 &0&\Omega\kappa&0&0&v&0&0\\
     0 &0&0& \frac{\kappa}{\Omega}&-v&0&0&0\\
     0&0&0&-v&\Omega\kappa&0&0&0\\
     0&0&v&0& 0&\frac{\kappa}{\Omega}&0&0\\
     0&0&0&0& 0&0&  \frac{1}{2}&0\\
     0&0&0&0&0 &0&0&   \frac{1}{2}\\
     \label{eq:Vi}
   \end{pmatrix}.
\eea
After the two beam splitters, the covariance matrix $\mVb_f = \mSb_{BS}(\frac{\pi}{2},\frac{\pi}{2})\mVb_i\mSb^\tp_{BS}(\frac{\pi}{2},\frac{\pi}{2})$ can be written in the block form
\begin{equation}
\mVb_f = \begin{pmatrix} A & B \\ B^\tp & A' \end{pmatrix},
\label{eq:Vfblock}
\end{equation}
where, defining 

\bse
\label{eq:defalfgam}
\begin{align}
\alpha_\pm &= \frac{1}{4}(2\kappa\Omega \pm 1), \\
\gamma_\pm &= \frac{2\kappa \pm \Omega}{4\Omega},
\end{align}
\ese
we have
\bse
\begin{align}
A &= \begin{pmatrix} \alpha_+ & 0 & \alpha_- & 0 \\ 0 & \gamma_+ & 0 & \gamma_{-}\\ \alpha_- & 0 & \alpha_+ & 0
\\ 0 & \gamma_{-} & 0 & \gamma_+ \end{pmatrix}, \label{eq:defMatA}\\
B &= \begin{pmatrix} 0 & \tfrac{v}{2} & 0 & -\tfrac{v}{2} \\ -\tfrac{v}{2} & 0 & \tfrac{v}{2} & 0 
\\ 0 & \tfrac{v}{2} & 0 & -\tfrac{v}{2} \\ -\tfrac{v}{2} & 0 & \tfrac{v}{2} & 0 \end{pmatrix}, \label{eq:defMatB}\\
A' &= \begin{pmatrix} \alpha_+ & 0 & -\alpha_- & 0 \\ 0 & \gamma_+ & 0 & -\gamma_- \\ -\alpha_- & 0 & \alpha_+ & 0 \\ 0 & -\gamma_- & 0 & \gamma_+ \end{pmatrix}.
\label{eq:defMatAp}
\end{align}
\ese
The off-diagonal block $B$ carries all dependence on $v$. Note that setting $v=0$ renders $\mVb_f$ block-diagonal.
\par

\subsection{Entanglement in 2 vs.  bipartitions of the form $WX|YZ$} 

There are three bipartitions with two modes in each part: $12|34$, $14|23$ and $13|24$.
\par
\noindent\textit{Bipartition $12|34$.}
Of course, no entanglement can be created in bipartition $12|34$, since with respect to this separation of modes, the beam splitters are purely local operations, acting on a separable four-mode state with CM given in Eq.~\eqref{eq:Vi}.
Because $\mVb_f$ is separable in the $12|34$ bipartition, the four-mode state is {biseparable} with respect to this cut, and be written as a mixture of product states $\hat{\rho}_{12}\otimes\hat{\rho}_{34}$~\cite{toscano2015}. A direct consequence is that no genuine tripartite or quadripartite entanglement can exist among all four modes simultaneously, and thus any entanglement present is necessarily bipartite in character, residing in bipartitions that cross the $12|34$ boundary.
\par
\medskip
\noindent\textit{Bipartition $14|23$.}
In order to test entanglement we apply the PPT criterion \cite{Simon2000} by computing the symplectic eigenvalues of the matrix
$\mVb_{f^{\tp(14)}}=\Lambda_{14}\mVb_{f}\Lambda_{14}$, where $\Lambda_{14}=\diag(1,-1,1,1,1,1,1,-1)$ corresponds to the partial transposition of modes $1$ and $4$.
The symplectic eigenvalues are the moduli of the eigenvalues of $\imath \mVb_{f^{\tp(14)}}\mJb$, with $\mJb$ given in Eq. \eqref{eq:J}. 
Although PPT is generally only necessary for separability across a
$2\times2$ Gaussian cut, the four-mode state $\mVb_f$ is bisymmetric under
$14|23$, being invariant under $1\leftrightarrow4$ and $2\leftrightarrow3$.
Such states are locally symplectically equivalent to a two-mode Gaussian state
tensor uncorrelated local modes~\cite{Serafini2005}, and for
two-mode Gaussian states PPT is necessary and sufficient for
separability~\cite{Simon2000}. Therefore, $f_{\min}<\tfrac12$ is both necessary
and sufficient for entanglement, and the conditions derived below
define exactly the separable region.
\par
There are only two distinct symplectic eigenvalues of $\mVb_{f^{\tp(14)}}$ (each is two-fold degenerate):
\begin{equation} f_\pm =
\sqrt{\frac{\lambda_{\pm}}{2}},
\label{eq:fabse1}
\end{equation}
with $\lambda_{\pm}$ the eigenvalues of the TGS state given in Eq. \eqref{eq:lammin}.
Note that $f_{min}=\sqrt{\frac{\lambda_{min}}{2}}<\frac{1}{2}\Leftrightarrow \lambda_{min}<\frac{1}{2}.$
 
\begin{figure}[t] 
    \centering
    \includegraphics[width=7cm]{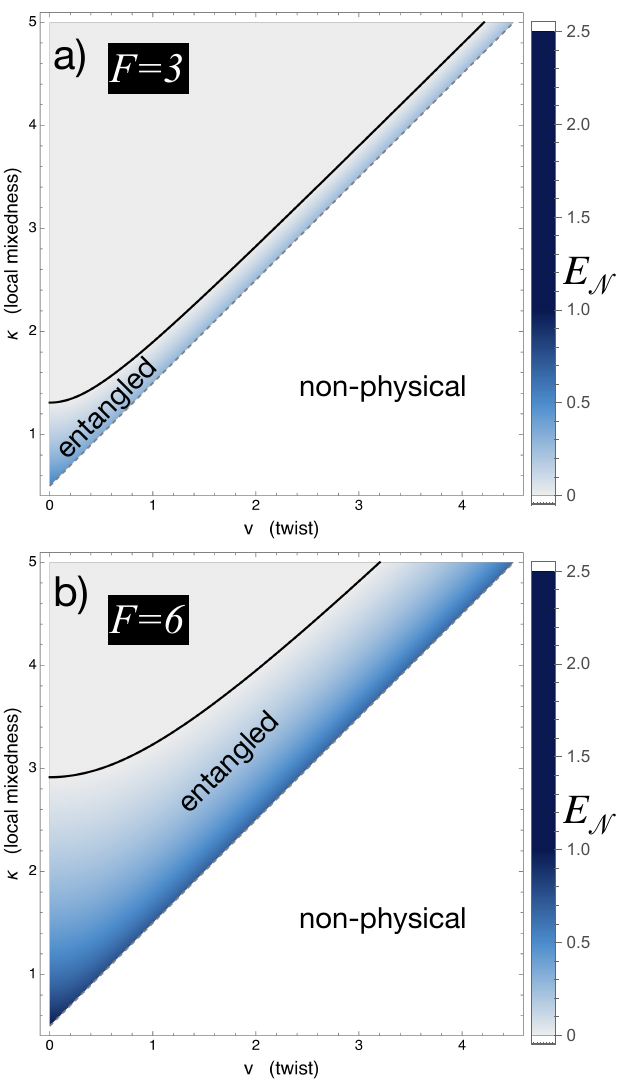}
    \caption{Entangled regions of the $(\kappa,v)$ plane of the $14|23$ bipartition for a) $F=3$ and b) $F=6$. The
dashed diagonal line is the physicality bound $v=\kappa-\tfrac12$. The solid black curve is
the entanglement boundary $\kappa F=T$ of Eq.~\eqref{eq:condentanglement}. Since $f_-<\tfrac12\Leftrightarrow\lambda_{\min}<\tfrac12$, this boundary is identical to the nonclassicality threshold $\lambda_{\min}=\tfrac12$ of the input TGS state (see Eq.~\eqref{eq:FlessFstar}). At fixed thermal noise
$\kappa$ and squeezing $\Omega$ ($F$), increasing the twist $v$ drives the state across the boundary. The shaded blue scale corresponds to the logarithmic negativity $E_\mathcal{N}$, with darker shade indicating larger $E_\mathcal{N}$. }
    \label{fig:kF-Tplots}
\end{figure}
Consequently, there is entanglement 
in the bipartition $14|23$ if the condition:
\begin{align}
\kappa F=\kappa\Omega+\frac{\kappa}{\Omega}> \frac{1}{2}\left(1+\frac{1}{\beta}\right)=T,
\label{eq:condentanglement}
\end{align}
is satisfied, that it is exactly the condition for squeezing in Eq.~\eqref{eq:FlessFstar}. 
\par
Eq.~\eqref{eq:condentanglement} establishes that entanglement is present in the bipartition $14|23$ if the quantity $\kappa F$ is greater than the threshold $T$ given in terms of the purity of the TGS state at the input, which has a direct physical meaning: the beam-splitter embedding
 activates entanglement in the $14|23$ bipartition if and only if the TGS
input is nonclassical, in accordance with the general result that passive linear
optics can entangle only nonclassical inputs~\cite{Kim2002,Asboth2005,Wolf2003}.
\par
 The two degenerate copies of $f_-$ combine so that the four-mode scheme attains exactly the same optimal logarithmic negativity as the minimal two-mode scheme (see Eq.~\eqref{eq:EN2mode}):
\begin{align}
E_\mathcal{N}&=\max\!\left(0,\,-2\log_2 2f_-\right)\nonumber\\
&=\max\!\left(0,\,-\log_2 2\lambda_{\min}\right).
\label{eq:ENfourmode}
\end{align}

It is worth noting that for fixed values of thermal noise $\kappa$ of the input TGS state, the threshold $T$ increases with the degree of correlations $0\leq v \leq \kappa-\frac{1}{2}$, achieving its maximum when $v=\kappa-\frac{1}{2}$, where Eq.~\eqref{eq:condentanglement} reduces to $F>2$, so that entanglement in bipartition $14|23$ is equivalent to the presence of any amount of squeezing in the TGS input state. 
\par
Interestingly, for fixed values of $\kappa$ and $\Omega$, an activation of entanglement can occur as $v$ increases. This can be seen in Fig.~\ref{fig:kF-Tplots}, where the solid curve marks the boundary $\kappa F = T$ of Eq.~\eqref{eq:condentanglement}, separating the separable region ($\kappa F<T$) from the entangled one ($\kappa F>T$) in the $(\kappa,v)$ plane, and the color encodes the logarithmic negativity of Eq.~\eqref{eq:ENfourmode}. Indeed, for certain fixed values of the 
thermal noise $\kappa$ and squeezing $F$, a transition from no entanglement to entanglement 
takes place as the correlations $v$ increase.
\par
Furthermore, within the region of parameter space that is already entangled at $v=0$,
increasing $v$ strictly increases the entanglement.
This follows from Eq.~\eqref{eq:fabse1}: the term $4v^2$ under the inner square root
is strictly positive for $v\neq 0$, which decreases $f_-$,
and therefore increases the logarithmic negativity.
The increment due to the twist can be given in closed form:
\begin{align}
\Delta\mathcal{E}_\mathcal{N}(v) &\equiv E_\mathcal{N}(v) - E_\mathcal{N}(v=0) \nonumber \\
&= \log_2\!\left(
\frac{\kappa\bigl(F - \sqrt{F^2-4}\bigr)}
{\kappa F - \sqrt{4v^2+\kappa^2(F^2-4)}}
\right),
\label{eq:DeltaEN}
\end{align}
valid throughout the region entangled at $v=0$, i.e., where
$\kappa(F - \sqrt{F^2-4}) < 1$.
This expression vanishes at $v=0$ and is manifestly positive for $v\neq 0$.
Both effects follow from the single fact that $v$ lowers $\lambda_{\min}$ (Sec.~\ref{sec:nonclassicality}): it enlarges the entangled region through the shift of the threshold $T$, and it increases the entanglement where the state is already entangled.
\par
\noindent\textit{Bipartitions $13|24$.} 
For this bipartition we have exactly the same result as for the bipartition $14|23$.


\subsection{Reduced two-mode states and the $14|23$ structure}
\label{sec:modepairs}

\par
The mode pairs $1,4$ and modes $2,3$, lie on opposite sides of the
separable $12|34$ cut, so the reduced states $\hrho_{14}$ and $\hrho_{23}$ are
themselves separable. They are nonetheless of particular interest, because rewriting
$\mVb_f$ in $14|23$ mode order exposes the structure that the four-mode embedding
preserves. Introducing the covariance matrix TGS-form template
\begin{equation}
\mathsf{T}(a,b;t)=\begin{pmatrix}
a&0&0&t\\ 0&b&-t&0\\ 0&-t&a&0\\ t&0&0&b
\end{pmatrix},
\label{eq:Ttemplate}
\end{equation}
the four-mode covariance matrix in $14|23$ order reads
\begin{equation}
\mVb_f^{(14|23)}=\begin{pmatrix}\mathcal A & \mathcal C\\
\mathcal C^{\tp}&\mathcal B\end{pmatrix},
\label{eq:Vf1423}
\end{equation}
with local two-mode blocks
$\mathcal A=\mathsf{T}\!\left(\alpha_+,\gamma_+;-\tfrac{v}{2}\right)$ and 
$\mathcal B=\mathsf{T}\!\left(\alpha_+,\gamma_+;+\tfrac{v}{2}\right)$,
and cross block
\begin{equation}
\mathcal C=\begin{pmatrix}
\alpha_- & 0 & 0 & v/2\\
0 & \gamma_- & -v/2 & 0\\
0 & v/2 & -\alpha_- & 0\\
-v/2 & 0 & 0 & -\gamma_-
\end{pmatrix},
\label{eq:Ccross}
\end{equation}

where $\alpha_{\pm}$ and $\gamma_{\pm}$ are defined in Eqs.~\eqref{eq:defalfgam}. Thus, the reduced states $\hrho_{14}$ and $\hrho_{23}$
correspond to TGS states whose 
covariance matrices are $\mathcal A$ and $\mathcal B$ respectively, 
 with effective parameters
$
\kappa_{\!A}=\kappa_{\!B}=\sqrt{\alpha_+\gamma_+}$,
$\Omega_{\!A}=\Omega_{\!B}=\sqrt{\alpha_+/\gamma_+}$,
$v_{\!A}=\tfrac{v}{2}$ and $v_{\!B}=-\tfrac{v}{2}$.

The embedding scheme therefore preserves the twist characteristic as the genuine TGS structure appear in its marginals.
 The cross block $\mathcal C$ carries the inter-mode couplings that entangle the two halves of the bipartition $14|23$, and also depends on $v$. Because the $14$ and $23$ marginals are themselves TGS states, they carry the same photon-number signature as the input (Sec.~\ref{sec:fockbasis}).
\par
Mode pairs on the same side of the $12|34$ cut, such as modes $1,2$, have a
reduced covariance matrix $\mVb_{12}=A$ [Eq.~\eqref{eq:defMatA}] independent of $v$;
their entanglement is the familiar two-mode type from splitting a single squeezed mode
at a beam splitter, present only when the input mode is squeezed below shot noise.

\subsection{Entanglement in bipartitions of the form $W|XYZ$} 
\label{Sec:entanglementW-XYZ}

There are four bipartitions of this type: $1|234$, $2|134$, $3|124$, and $4|123$.
In the analysis of entanglement for these four bipartitions, the same equations arise, and the results are therefore completely equivalent, reflecting the symmetry of the beam-splitter construction. We thus present only the result for the bipartition $1|234$.
\par
It is worth noting that for Gaussian states in bipartitions of the form $1|(n-1)$, where $n$ is the total number of bosonic modes, the PPT criterion provides a necessary and sufficient condition for entanglement~\cite{Werner2001}.
In Appendix~\ref{App:entaW-XYZ}, we derive the conditions under which the smallest symplectic eigenvalue of the partially transposed matrix $\mVb_{f^{\tp(1)}}$ drops below the threshold $\frac{1}{2}$ to detect entanglement in the bipartition. 
Indeed, we prove that this coincides precisely with the entanglement condition~\eqref{eq:condentanglement} found for the bipartitions $14|23$ and $13|24$. Hence the four bipartitions $W|XYZ$ are entangled over exactly the same region of parameter space as the bipartitions $14|23$ and $13|24$, and the twist $v$ plays the same role.

\section{Conclusion}
\par
We have introduced the Twisted Gaussian Schell (TGS) state, a family of two-mode mixed Gaussian states defined as the quantum-optical analog of the classical Twisted Gaussian Schell-model beam. We have  establishing an explicit dictionary between classical beam parameters (beam width $\omega_0$, coherence length $\delta_0$, twist phase $\mu_0$) and the quantum parameters $(\kappa, \Omega, v)$ of the covariance matrix. We show that the state can be generated from a two-mode thermal state by local squeezing between passive linear-optical networks, providing a concrete experimental recipe, and the classical physicality bound on the twist coincides exactly with the quantum physicality condition. The twist parameter $v$ leaves a direct signature in the joint photon-number distribution: the twist-induced coupling prevents $P(n_1,n_2)$ from factorizing and produces nonzero mutual information despite separability. 
\par
The central quantum feature of the TGS state is its nonclassicality. Although separable in its natural bipartition, the state can be squeezed below the shot-noise limit.  For non-zero twist parameter the squeezed quadrature is defined in the two modes, and the twist lowers the smallest quadrature variance, acting as a resource that drives the state below threshold once any local squeezing is present. 
\par
The non-classicality allows for entanglement under the action of passive optics. The attainable logarithmic negativity is a monotone function of the smallest variance alone, and we consider two mechanisms to produce entanglement. The first, a phase shift and balanced beam splitter, entangle the two TGS modes optimally but transform the twist away. A second scheme involves coupling each mode to an ancillary vacuum, which reaches the same logarithmic negativity while keeping the twist intact.
Thus,
across the $14|23$ or $13|24$ bipartitions, the four-mode state is an entangled pair locally described by TGS states, while $12|34$ bipartition remains separable. Entanglement is also present in all $W|XYZ$ splits. Since Gaussian entanglement depends only on the covariance matrix, replacing the vacuum ancillas with coherent states leaves these results unchanged and turns the network into a double homodyne setup.
\par

\par
These results open several directions for future work. On the theoretical side, it would be interesting to explore whether other partially coherent beams, when translated into a two-mode quantum formalism, give rise to states with analogous or richer entanglement structures. On the experimental side, the generation scheme derived here, combined with homodyne tomography or photon-counting measurements, suggests a pathway to verify twist-induced Fock-basis correlations and twist-assisted entanglement in the laboratory.

\begin{acknowledgments}
This research was funded by Chilean Fondo Nacional de Desarrollo
Cient\'{\i}fico y Tecnol\'ogico (FONDECYT) Grant Nos.\ 1240746 and 1230796,
ANID -- Millennium Science Initiative Program -- ICN17$\_$012,
ANID Anillo Project ATE250003, and by the following Brazilian research agencies: Conselho Nacional de Desenvolvimento Cient\'{\i}fico e Tecnol\'ogico (CNPq - DOI 501100003593), Coordena\c c\~{a}o de Aperfei\c coamento de Pessoal de N\'\i vel Superior (CAPES DOI 501100002322), Funda\c c\~{a}o de Amparo \`{a} Pesquisa do Estado de Santa Catarina (FAPESC - DOI 501100005667), Instituto Nacional de Ci\^encia e Tecnologia de Informa\c c\~ao Qu\^antica (INCT-IQ 465469/2014-0, INCT-IQNano 406636/2022-2 and INCT-DQ 408783/2024-9),     Funda\c c\~ao Carlos Chagas Filho de Amparo à Pesquisa do Estado do Rio de Janeiro (FAPERJ Process E-26/203.939/2024), Funda\c c\~ao de Amparo à Pesquisa do Estado de São Paulo (FAPESP Process 2021/06823-5).  
\end{acknowledgments}
\appendix

\section{The Euler decomposition of $\mSb$ in Eq.~\eqref{eq:defSmb}}
\label{AppEulerDecomp}

The Euler decomposition $\mSb = \mOb\mSb_d\mObp=\mSigb\mOmb$ is obtained via the polar decomposition
\begin{subequations}
\begin{align}
\mSigb &= (\mSb\mSb^\tp)^{1/2}=\mOb\mSb_d\mOb^\tp, \\
\mOmb  &= (\mSb\mSb^\tp)^{-1/2}\mSb = \mOb\mObp,
\end{align}
\end{subequations}
where $\mSigb$ is symplectic and positive definite, and $\mOmb$ is symplectic and orthogonal. {Direct computation of $\mSb\mSb^\tp$ from Eq.~\eqref{eq:defSmb} gives
\begin{equation}
\mSb\mSb^\tp = \mOb \begin{pmatrix} \Omega^{-1} &0 & 0& 0\\0 & \Omega & 0& 0\\ 0&0 & \Omega^{-1} & 0\\ 0& 0& 0& \Omega \end{pmatrix} \mOb^\tp,
\end{equation}}
where $\mOb$ is the orthogonal symplectic matrix
\begin{equation}
\mOb = \frac{1}{\sqrt{2}}\begin{pmatrix} 0 & -1 & 1 & 0 \\ 1 & 0 & 0 & 1 \\ 0 & 1 & 1 & 0 \\ -1 & 0 & 0 & 1 \end{pmatrix}.
\end{equation}
Taking the positive square root and inverse yields the diagonal squeezing matrix
\begin{equation}
\mSb_d = \diag\!\left(\tfrac{1}{\sqrt{\Omega}},\,\sqrt{\Omega},\,\tfrac{1}{\sqrt{\Omega}},\,\sqrt{\Omega}\right),
\end{equation}
and the orthogonal factor
\begin{equation}
\mObp = \mOb^\tp \mOmb = \begin{pmatrix} -1 & 0 & 0 & 0 \\ 0 & -1 & 0 & 0 \\ 0 & 0 & 1 & 0 \\ 0 & 0 & 0 & 1 \end{pmatrix}.
\end{equation}
The orthogonal matrix $\mOb$ decomposes as
\begin{equation}
\mOb=\mS_{FE}(-\tfrac{\pi}{2},-\tfrac{\pi}{2})\,\mS_{BS}(\tfrac{\pi}{2})\,\mS_{PS}(\tfrac{\pi}{2}),
\end{equation}
{where the symplectic matrices for a beam splitter, a phase shifter, and free evolution are:
\begin{align}
\mS_{BS}(\phi)= \begin{pmatrix} 
c_\phi& 0&s_\phi&0 \\
      0 & c_\phi &0&s_\phi\\
     -s_\phi&0& c_\phi&0\\
    0 &-s_\phi&0& c_\phi
   \end{pmatrix},
\end{align}
\begin{align}
\mS_{PS}(\phi) &= \begin{pmatrix} c_\phi & -s_\phi & 0 & 0 \\ s_\phi & c_\phi & 0 & 0 \\ 0 & 0 & c_\phi & s_\phi \\ 0 & 0 & -s_\phi & c_\phi \end{pmatrix}, \\[4pt]
\mS_{FE}(\psi,\theta) &= \begin{pmatrix} c_\psi & s_\psi & 0 & 0 \\ -s_\psi & c_\psi & 0 & 0 \\ 0 & 0 & c_\theta & s_\theta \\ 0 & 0 & -s_\theta & c_\theta \end{pmatrix},
\end{align}
where $c_x = \cos(x/2)$ and $s_x = \sin(x/2)$ for $x=\phi,\psi,\theta$. 
Noting that $\mS_{FE}(2\pi,0) = \diag(-1,-1,1,1) = \mObp$, the full Euler decomposition is
\begin{align}
\mSb &= \mS_{FE}(-\tfrac{\pi}{2},-\tfrac{\pi}{2})\,\mS_{BS}(\tfrac{\pi}{2})\,\mS_{PS}(\tfrac{\pi}{2}) \nonumber\\
&\quad\times\diag\!\left(\tfrac{1}{\sqrt{\Omega}},\sqrt{\Omega},\tfrac{1}{\sqrt{\Omega}},\sqrt{\Omega}\right)
\mS_{FE}(2\pi,0),
\end{align}
in agreement with Eq.~\eqref{eq:euler}.

\section{Derivation of joint photon number distribution of TGS state}
\label{App:derPn1n2}
To calculate $P(n_1,n_2)$ from Eq. \eqref{eq:Pn1n2ini}, we use~\footnote{The factor $2^{2n}$ is missing in the formula from page 251 of the book~\cite{magnus2013formulas}, but using Mathematica it seems that it is there.}:
\begin{align}
L_n(x^2+y^2)&=\frac{(-1)^n}{2^{2n}n!}\sum_{m=0}^{n}\binom{n}{m}\nonumber\\
&\times H_{2m}(x)H_{2n-2m}(y),
\label{eq:LninterHn}
\end{align}
where the Hermite polynomials have the generating function~\cite{magnus2013formulas}:
\begin{align}
\sum_{n=0}^{+\infty}\frac{1}{n!} H_n(x) y^n=e^{-y^2+2yx}.
\label{eq:genfunherpoly}
\end{align} 
Replacing Eq.~\eqref{eq:LninterHn} into~\eqref{eq:Pn1n2ini} we obtain
\begin{align}
P(n_1,n_2)&=\frac{C}{(\pi)^2}
\frac{1}{2^{2(n_1+n_2)}n_1! n_2!}
\sum_{k_1=0}^{n_1}\sum_{k_2=0}^{n_2}
 \binom{n_1}{k_1} \binom{n_2}{k_2} \nonumber\\
 &\times 
\left( \int dp_1 \;e^{-bp_1^2} \, H_{2(n_1-k_1)}\left(\sqrt{2}p_1\right) \,\calI_{2k_2}(p_1)\right)\nonumber\\
&\times \left( \int dp_2\;e^{-bp_2^2} \, H_{2(n_2-k_2)}\left(\sqrt{2}p_2\right)\,\calI_{2k_1}(p_2)\right),
\end{align}
where we define the integral \[\calI_{2k}(p)=\int_{-\infty}^{\infty}\;dx\;e^{-a\,x^2\pm c\,x\,p}\;H_{2k}(\sqrt{2}x)\]
and the parameter $\gamma=\left(1-\frac{2}{a}\right)^{1/2}$.
\par
Using the result proved in Appendix~\ref{App:usefulinte}, \ie
\begin{align}
\calI_{2k}(p)&=\sqrt{\frac{\pi}{a}}e^{\frac{c^2p^2}{4a}}\gamma^{2k}
H_{2k}\left(\frac{cp}{\sqrt{2}a\gamma}\right),
\label{eq:usetwointe1}
\end{align}
we arrive at 
\begin{align}
P(n_1,n_2)&=\frac{C}{\pi \,a}
\frac{1}{2^{2(n_1+n_2)}n_1! n_2!}
\sum_{k_1=0}^{n_1}\sum_{k_2=0}^{n_2}\binom{n_1}{k_1} \binom{n_2}{k_2}\nonumber\\
&\times  \gamma^{2(k_1+k_2)}
 \calI_{2(n_1-k_1)\,2k_2} \left(\eta,\sqrt{2},\delta\right)\nonumber\\
&\times  \calI_{2(n_2-k_2)\,2k_1} \left(\eta,\sqrt{2},\delta\right),
\label{eq:Pn1n2provi} 
\end{align}
where we define the integral
$\calI_{2m\,2n}(\eta,\alpha,\delta) =\int_{-\infty}^{\infty} e^{-\eta x^2} H_{2m}(\alpha x) H_{2n}(\delta x) \,dx$,
and the parameter
$\delta=\frac{c}{\sqrt{2}a\gamma}$.
\par
Replacing the result proved in Appendix~\ref{App:usefulinte2}:
\begin{align}
\calI_{2m\,2n}(\eta,\alpha,\delta)&= \sqrt{\pi}\eta^{-(m+n)-\frac{1}{2}}
(\alpha^2-\eta)^{m}(\delta^2-\eta)^{n}\nonumber\\
&\times \sum_{j=0}^{\min(m,n)} 
\frac{(2m)!\,(2n)!}{(m-j)!\,(n-j)!\,(2j)!}\nonumber\\
&\times\left(\frac{4\alpha^2\delta^2}{(\alpha^2-\eta)(\delta^2-\eta)}\right)^j,
\label{eq:usetwointe2}
\end{align}
into~\eqref{eq:Pn1n2provi} and rearranging terms we arrive to Eq.~\eqref{eq:finalPn1n2}.

\section{Derivation of marginal photon number distributions}
\label{App:derPnj}

Because the expression in Eq.~\eqref{eq:Pn1n2ini} is symmetric with respect to $n_1$ and $n_2$, it suffices to compute the marginal distribution: $P(n_1) = \sum_{n_2=0}^{\infty} P(n_1,n_2)$.
Summing over $n_2$ and using 
\beq
\sum_{n_2=0}^{\infty} (-1)^{n_2} L_{n_2}(2(x_2^2+p_2^2)) = \frac{1}{2} e^{x_2^2 + p_2^2},
\eeq
we obtain:
\begin{align}
P(n_1) 
&= \frac{(-1)^{n_1}}{2\pi^2}\,C \int \int dx_1\,dp_1\,e^{-a x_1^2 - b p_1^2}\nonumber\\
&\times L_{n_1}(2(x_1^2+p_1^2))\nonumber\\
&\times \left(\int dx_2\,
e^{-(a-1)x_2^2 - c p_1 x_2}\right) \nonumber\\
&\times \left(\int  dp_2 \,
e^{- (b-1)p_2^2 + c x_1 p_2 }\right).
\end{align}
Note that because of~\eqref{eq:cona} we can perform the Gaussian integrals, \ie $\int_{-\infty}^{\infty} dx_2\, e^{-(a-1)x_2^2 - c p_1 x_2} = \sqrt{\frac{\pi}{a-1}}\, \exp\left( \frac{c^2 p_1^2}{4(a-1)} \right)$ and $\int_{-\infty}^{\infty} dp_2\, e^{-(b-1)p_2^2 + c x_1 p_2} = \sqrt{\frac{\pi}{b-1}}\, \exp\left( \frac{c^2 x_1^2}{4(b-1)} \right)$, to obtain,
\begin{align}
P(n_1) &= \frac{(-1)^{n_1} C}{2\pi \sqrt{(a-1)(b-1)}} 
\int \int dx_1\,dp_1\,
e^{-\chi x_1^2} 
e^{- \xi p_1^2 }\nonumber\\
&\times L_{n_1}(2(x_1^2+p_1^2)),
\label{eq:Pn1cominte}
\end{align}
where we define the parameters 
\bse
\label{eq:defchixi}
\begin{align}
\chi&=a-\frac{c^2}{4(b-1)},\\
\xi&=b-\frac{c^2}{4(a-1)}. 
\end{align}
\ese
\par
Now using Eq.~\eqref{eq:LninterHn} into~\eqref{eq:Pn1cominte}
we arrive at
\begin{align}
P(n_1) &= \frac{(-1)^{n_1} C}{2\pi \sqrt{(a-1)(b-1)}} 
\frac{(-1)^{n_1}}{2^{2n_1}n_1!} \sum_{k=0}^{n_1} \binom{n_1}{k} \nonumber\\
&\times \left(\int dx_1\,e^{-\chi x_1^2  }
H_{2k}(\sqrt{2}x_1)\right)\nonumber\\
&\times \left(
\int \,dp_1\,e^{- \xi p_1^2 }
H_{2(n_1-k)}(\sqrt{2}p_1)\right).
\label{eq:Pn1doisinte}
\end{align}
Notice that the positive constant
in~\eqref{eq:defC} can be written as $C=4(a-1)(b-1)-c^2$. So, from $C>0$ we immediately obtain that $\chi,\xi>1$, what guarantees the convergence of the integrals in~\eqref{eq:Pn1doisinte}.
\par
Using the result in~\eqref{eq:usetwointe1} to perform the integrations in~\eqref{eq:Pn1doisinte}, \ie 
$\int_{-\infty}^{\infty} e^{- o\,x_1^2} H_{2m}(\sqrt{2} x) \, dx = 
\sqrt{\frac{\pi}{o}} \, \frac{(2m)!}{m!} \left( \frac{2}{o} - 1 \right)^m$, we arrive at
\begin{align}
P(n_1) &=
\frac{C}{2\sqrt{(a-1)(b-1)\chi\xi}} \frac{1}{2^{2n_1}}\sum_{k=0}^{n_1} 
\binom{2k}{k}\nonumber\\
&\times \binom{2(n_1 - k)}{n_1 - k} 
u^k v^{(n_1-k)},
\end{align}
where $u=\frac{2}{\chi}-1$ and $w=\frac{2}{\xi}-1$ are exactly the parameters defined in Eqs.~\eqref{eq:defu} and~\eqref{eq:defw}, once we use the definitions in Eqs.~\eqref{eq:defchixi}. Finally we obtain the expression in Eq.~\eqref{eq:formulaforPnj} using the identity, which follows from the generating functions of the central binomial coefficients and the Legendre polynomials~\cite{Lebedev1965},
\begin{align}
\sum_{k=0}^{n} \binom{2k}{k} \binom{2(n-k)}{n-k} u^k w^{n-k}
&=2^{2n} (\sqrt{uw})^n \nonumber\\
&\times P_n\!\left( \frac{u+w}{2\sqrt{uw}} \right).
\label{eq:convcentbicoeff}
\end{align}

\section{Proof of the integral in Eq.\eqref{eq:usetwointe1}}
\label{App:usefulinte}
\par
Using the generating function of the Hermite polynomials in~\eqref{eq:genfunherpoly} and performing the Gaussian integral, we have
\begin{align}
\sum_{q=0}^{\infty} \frac{t^{q}}{q!} \calI_q(p)&=
e^{-t^2} \int_{-\infty}^{\infty}\;dx\;e^{-ax^2+\left(2\sqrt{2}t\pm cp\right)x} \nonumber\\
&=
\sqrt{\frac{\pi}{a}} e^{\frac{c^2 p^2}{4 a}}
\sum_{q,s\geq 0}(-1)^s\frac{\left(\pm\frac{\sqrt{2}cp}{a}\right)^q}{q!}\nonumber\\
&\times \frac{\gamma^{2s}}{s!} t^{q+2s},
\label{eq:auxi1}
\end{align}
with $\gamma^2=1-\frac{2}{a}$. Matching the coefficient of $t^{2k}$ on both sides (so $q+2s=2k$, with $q=2j$ and $s=k-j$, $0\leq j \leq k$) and using $H_{2k}(z) = (2k)! \sum_{j=0}^{k} \frac{(-1)^{k-j} (2z)^{2j}}{(k-j)! \, (2j)!}$ gives
\begin{align}
\calI_{2k}(p)
&=
\sqrt{\frac{\pi}{a}} e^{\frac{c^2 p^2}{4 a}}\gamma^{2k}\,H_{2k}\left(\frac{cp}{\sqrt{2}a\gamma}\right).
\end{align}

\section{Proof of the integral in Eq.\eqref{eq:usetwointe2}}
\label{App:usefulinte2}

Here we prove the identity in Eq.\eqref{eq:usetwointe2} that it is valid for
$\operatorname{Re}(\eta)>0$ and arbitrary complex numbers $\alpha,\delta$.
Using the generating function of the Hermite polynomials in~\eqref{eq:genfunherpoly}
for two independent variables $t,u$, we have,
\begin{align*}
\sum_{p,q=0}^{\infty} \frac{t^{p}u^{q}}{p!\,q!} H_p(\alpha x) H_q(\delta x)
= e^{2\alpha x t - t^2}\; e^{2\delta x u - u^2}.
\end{align*}
Now, multiplying by $e^{-\eta x^2}$, using the definition in Eq~\eqref{eq:defIpq}, and integrating over $x$ we get
\begin{align}
\sum_{p,q=0}^{\infty} \frac{t^{p}u^{q}}{p!\,q!}
I_{pq}
&= e^{-(t^2+u^2)}\int_{-\infty}^{\infty} e^{-\eta x^2 + 2(\alpha t+\delta u)x}dx
\label{eq:sumIpq}\\
&=\sqrt{\frac{\pi}{\eta}}\exp\left(F t^2+Gu^2 +Htu\right)\label{eq:sumIpq3}\\
&= \sqrt{\frac{\pi}{\eta}}\sum_{p,q,r\ge0} \frac{F^p}{p!}\frac{G^q}{q!}\frac{H^r}{r!}\, t^{2p+r} u^{2q+r}.
\label{eq:sumIpq4}
\end{align}
In the left-hand side (lhs) of~\eqref{eq:sumIpq} we defined
\beq
\label{eq:defIpq}
\calI_{pq}(\eta,\alpha,\delta)=\int_{-\infty}^{\infty} e^{-\eta x^2} H_{p}(\alpha x) H_{q}(\delta x) \,dx,
\eeq
we have also performed the standard Gaussian integration in~\eqref{eq:sumIpq} and we expanded
the exponential in~\eqref{eq:sumIpq3} setting 
\bse
\label{eq:defABC}
\begin{align}
F &= \frac{\alpha^2}{\eta}-1,\\
G &= \frac{\delta^2}{\eta}-1,\\
H &= \frac{2\alpha\delta}{\eta}.
\end{align}
\ese
\par
Because we need the coefficient of $t^{2m}u^{2n}$ we require on the right-hand side (rhs)
of~\eqref{eq:sumIpq4} that
$2p+r = 2m, \,2q+r = 2n$.
Thus $r$ must be even so $r=2j$. Then $p = m-j$, $q = n-j$, with $0\le j\le\min(m,n)$.
Equating the coefficients in lhs of Eq.\eqref{eq:sumIpq} with the coefficients in Eq.\eqref{eq:sumIpq4}, and multiplying both sides by $(2m)!\,(2n)!$ gives
\begin{align}
I_{2m,2n}
&= \sqrt{\frac{\pi}{\eta}}\;
\sum_{j=0}^{\min(m,n)} \frac{(2m)!\,(2n)!}{(m-j)!\,(n-j)!\,(2j)!}\nonumber\\
&\times F^{m-j}G^{n-j}H^{2j}.
\label{eq:In1n2inter}
\end{align}
Replacing $F,G$ and $H$ from Eq.~\eqref{eq:defABC} and regrouping the powers of $\eta$ gives
\begin{align}
F^{m-j}G^{n-j}H^{2j}&=\eta^{-(m+n)}(\alpha^2-\eta)^{m}(\delta^2-\eta)^{n}\nonumber\\
&\times \left(\frac{4\alpha^2\delta^2}{(\alpha^2-\eta)(\delta^2-\eta)}\right)^{j}.
\end{align} 
Replacing in~\eqref{eq:In1n2inter} we recover exactly the formula in~\eqref{eq:usetwointe2}.

\section{Symplectic eigenvalues in 1 vs. 3 mode bipartitions}
\label{App:entaW-XYZ}

The symplectic eigenvalues of $\mVb_{f^{\tp(1)}}=\Lambda_{1}\mVb_{f}\Lambda_{1}$, where $\Lambda_{1}=\diag(1,-1,1,1,1,1,1,1)$ corresponds to the partial transposition of 
mode $1$, are the moduli of the solutions of the characteristic equation for the eigenvalues of the 
matrix $\imath \mVb_{f^{\tp(1)}}\mJb$, i.e.,
\begin{align}
\left(4 f ^2-1\right) 
\left( z^3 + a_2\,z^2 +a_1\,z +a_0\right)=0,
\label{eq:characeq}
\end{align}
where $z=f^2$, 
\begin{align}
a_2&=-\left(\kappa ^2+\frac{\kappa F}{2}\right),\\
a_1&=\kappa\left(\frac{F}{2}(\kappa^2-v^2)+\frac{\kappa }{4}\right),\\
a_0&=-\frac{1}{4}(\kappa^2-v^2)^2,
\end{align}
and $F$ is defined in Eq.~\eqref{eq:defF}.
\par
It is worth stressing that all solutions of the characteristic equation of the matrix
$\imath \mVb_{f^{\tp(1)}}\mJb$ are always real because $\mVb_{f^{\tp(1)}}$ is positive definite, being a congruence $\Lambda_1\mVb_f\Lambda_1$ of the bona fide covariance matrix $\mVb_f$ by the invertible $\Lambda_1$. Consequently, $\imath \mVb_{f^{\tp(1)}}\mJb$ has a purely real spectrum and satisfies Williamson's theorem~\cite{williamson}. In particular, every root $z_k=f^2$ of the cubic in Eq.~\eqref{eq:characeq} is real and non-negative.
\par
Two eigenvalues of $\imath \mVb_{f^{\tp(1)}}\mJb$ arise from $4 f ^2-1=0$ in Eq.~\eqref{eq:characeq}, so the corresponding symplectic eigenvalues of $\mVb_{f^{\tp(1)}}$ are $f=\{\frac{1}{2},\frac{1}{2}\}$.
The remaining six eigenvalues of $\imath \mVb_{f^{\tp(1)}}\mJb$ are positive, so 
the corresponding six symplectic eigenvalues of $\mVb_{f^{\tp(1)}}$ are
$f=\{\sqrt{z_0},\sqrt{z_0},\sqrt{z_1},\sqrt{z_1}, \sqrt{z_2},\sqrt{z_2}\}$, where $z_k$, with $k=0,1,2$, are the solutions of the cubic equation in~\eqref{eq:characeq}.
\par

The bipartition $1|234$ is entangled if and only if the smallest root $z_{\min}$ of the cubic $P(z)=z^3+a_2 z^2+a_1 z+a_0$ in Eq.~\eqref{eq:characeq} satisfies $z_{\min}<\frac{1}{4}$. As its roots are real and non-negative, $P$ has the standard shape: it rises from $P(0)=a_0=-\frac{1}{4}(\kappa^2-v^2)^2<0$, crosses its smallest root $z_0$, peaks, dips through $z_1$, and rises through $z_2$ (Fig.~\ref{fig:WXYZcubic}). Evaluating $P$ at the threshold $z=\frac{1}{4}$ gives the factorization
\begin{align}
\label{eq:Pquarter}
P\!\left(\tfrac{1}{4}\right)=\frac{1}{64}\bigl[4(\kappa^2-v^2)-1\bigr]\bigl[2\kappa F-4(\kappa^2-v^2)-1\bigr].
\end{align}
Physicality ($\kappa\ge\frac{1}{2}+v$, $v\ge 0$) gives $\kappa^2-v^2=(\kappa-v)(\kappa+v)\ge\frac{1}{4}$, so the first bracket is non-negative and the sign of $P(\frac{1}{4})$ is that of $2\kappa F-4(\kappa^2-v^2)-1$.

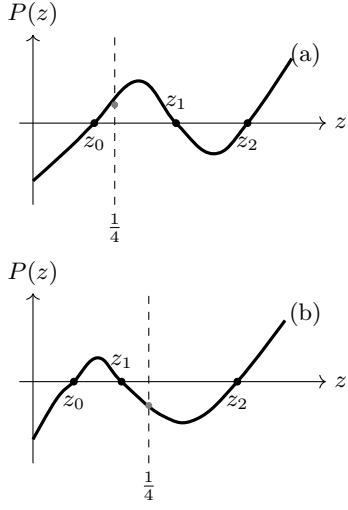
\begin{figure}[t]
\centering
\begin{tikzpicture}[scale=0.9]
\begin{scope}
\draw[->] (-0.2,0)--(4.3,0) node[right]{$z$};
\draw[->] (0,-1.2)--(0,1.3) node[above]{$P(z)$};
\draw[dashed] (1.2,-1.2)--(1.2,1.3);
\node[below] at (1.2,-1.2) {$\tfrac14$};
\draw[very thick] plot[smooth,tension=0.7] coordinates
  {(0,-0.85) (0.55,-0.35) (0.9,0) (1.55,0.62) (2.1,0) (2.65,-0.45) (3.15,0) (3.8,0.95)};
\filldraw (0.9,0) circle (1.4pt) node[below=2pt]{$z_0$};
\filldraw (2.1,0) circle (1.4pt) node[above=2pt]{$z_1$};
\filldraw (3.15,0) circle (1.4pt) node[below=2pt]{$z_2$};
\filldraw[gray] (1.2,0.27) circle (1.2pt);
\node[align=center] at (4.0,1.0) {(a)};
\end{scope}
\begin{scope}[yshift=-3.8cm]
\draw[->] (-0.2,0)--(4.3,0) node[right]{$z$};
\draw[->] (0,-1.2)--(0,1.3) node[above]{$P(z)$};
\draw[dashed] (1.7,-1.2)--(1.7,1.3);
\node[below] at (1.7,-1.2) {$\tfrac14$};
\draw[very thick] plot[smooth,tension=0.7] coordinates
  {(0,-0.85) (0.35,-0.25) (0.6,0) (0.95,0.35) (1.3,0) (1.9,-0.5) (2.4,-0.55) (3.0,0) (3.7,0.9)};
\filldraw (0.6,0) circle (1.4pt) node[below=2pt]{$z_0$};
\filldraw (1.3,0) circle (1.4pt) node[above=2pt]{$z_1$};
\filldraw (3.0,0) circle (1.4pt) node[below=2pt]{$z_2$};
\filldraw[gray] (1.7,-0.35) circle (1.2pt);
\node[align=center] at (4.0,1.0) {(b)};
\end{scope}
\end{tikzpicture}
\caption{The sign of $P(\tfrac14)$ decides entanglement, provided the situation in panel (b) cannot occur. Panel (a): the honest case, a single crossing before $\tfrac14$, with $P(\tfrac14)>0$. Panel (b): the ``double dip'', in which the curve rises above the axis and back below it before $\tfrac14$, hiding two roots there while $P(\tfrac14)<0$. The text shows panel (b) never happens, because the peak of $P$ cannot lie to the left of $\tfrac14$ when $P(\tfrac14)\le0$.}
\label{fig:WXYZcubic}
\end{figure}

We claim that $z_{\min}<\frac{1}{4}$ if and only if $P(\frac{1}{4})>0$. If $P(\frac{1}{4})>0$, then since $P(0)<0$ the cubic crosses zero in $(0,\frac{1}{4})$, so $z_{\min}<\frac{1}{4}$ [Fig.~\ref{fig:WXYZcubic}(a)]. For the converse we must exclude the only failure mode: that \emph{two} roots lie below $\frac{1}{4}$ while $P(\frac{1}{4})$ is again negative [the ``double dip'' of Fig.~\ref{fig:WXYZcubic}(b)]. A direct computation from $a_1$ and $a_2$ gives
\begin{equation}
\label{eq:Pprime}
P'\!\left(\tfrac{1}{4}\right)=\frac{3-4\kappa^2}{16}+\frac{\kappa F}{4}\,(2s-1),\qquad s\equiv\kappa^2-v^2,
\end{equation}
and we first show that $P(\frac{1}{4})\le 0$ implies $P'(\frac{1}{4})\ge 0$. Physicality gives $s\ge v+\frac{1}{4}\ge\frac{1}{4}$, $s\ge\kappa-\frac{1}{4}$ and $\kappa F\ge 2\kappa$. If $s\ge\frac{1}{2}$, then $2s-1\ge 0$ and, using $\kappa F\ge 2\kappa$ followed by $s\ge\kappa-\frac{1}{4}$,
\begin{equation}
P'\!\left(\tfrac{1}{4}\right)\ge\frac{3-4\kappa^2}{16}+\frac{\kappa}{2}\Bigl(2\kappa-\frac{3}{2}\Bigr)=\frac{3(2\kappa-1)^2}{16}\ge 0.
\end{equation}
If instead $\frac{1}{4}\le s<\frac{1}{2}$, then $2s-1<0$, so Eq.~\eqref{eq:Pprime} is decreasing in $\kappa F$; since $P(\frac{1}{4})\le 0$ is equivalent to $\kappa F\le 2s+\frac{1}{2}$,
\begin{align}
P'\!\left(\tfrac{1}{4}\right)\ge\frac{(4s-1)^2-4v^2}{16}\ge 0,
\end{align}
using $4s-1\ge 4v\ge 2v$. To turn this into a statement about the peak, note that $P'$ is an upward parabola, negative only on the interval $(c_1,c_2)$ between its roots, which are the peak $c_1$ and trough $c_2$ of $P$. Its midpoint is $\frac{1}{2}(c_1+c_2)=-\frac{a_2}{3}=\frac{\kappa^2}{3}+\frac{\kappa F}{6}\ge\frac{1}{4}$, so this interval lies at or to the right of $\frac{1}{4}$. Hence $P'(\frac{1}{4})\ge 0$ places $\frac{1}{4}$ outside the interval, and it can only be to the \emph{left}, $\frac{1}{4}\le c_1$ (it cannot be to the right, since that side begins beyond the midpoint, already $\ge\frac{1}{4}$). The peak therefore sits at or beyond $\frac{1}{4}$, so $P$ is nondecreasing on $[0,\frac{1}{4}]$ and, with $P(0)<0$, stays $\le 0$ there: the cubic has no root below $\frac{1}{4}$, i.e. $z_{\min}\ge\frac{1}{4}$. This excludes the double dip and establishes the equivalence. The bipartition $1|234$ is therefore entangled if and only if
\begin{equation}
\label{eq:condentWXYZapp}
\kappa F>\frac{1}{2}+2(\kappa^2-v^2)=\frac{1}{2}\left(1+\frac{1}{\beta}\right)=T,
\end{equation}
which is precisely the entanglement condition~\eqref{eq:condentanglement} obtained for the bipartition $14|23$. By the beam-splitter symmetry, the four $W|XYZ$ bipartitions are entangled over the same region of parameter space as $14|23$.

%

\end{document}